\documentstyle[12pt,epsfig]{article}

\newcommand{\eeql}[1]{\label{#1}\end{equation}}
\newcommand{\eeal}[1]{\label{#1}\end{eqnarray}}

\catcode`\@=11
\def\simless{\raise.3ex\hbox{$<$\kern-.75em\lower 1ex\hbox{$\sim$}}}
\def\grsim{\raise.3ex\hbox{$>$\kern-.75em\lower 1ex\hbox{$\sim$}}}

\def\Avalue{49}
\def\Cvalue{109}
\def\Cvaluecgs{$4.5\times10^{-19}\,$erg\thinspace cm}
\def\Ceffvalue{82}
\def\badC{85}
\def\Fvalue{-1.6} 

\def\Dvalue{67} 
\def\Nvalue{69}
\def\Lvalue{15.6}
\def\thresh{3}
\def\dd{{\rm d}}

\def\@maketitle{\newpage     
 \null
 \vspace*{-1\headsep}     
 \vspace*{-1\headheight}
 \vspace*{-24pt}
 \begin{flushright}{\absize
   { \preprintno} \\ \@date}
 \end{flushright}
 \vskip \headsep     
 \vskip \headheight
 \bigskip
 \begin{center}     
   {\tisize\bf \@title \par}
   \vskip 2em
   {\ausize
     \begin{tabular}[t]{c}\@author
     \end{tabular}\par}
   \vskip 1ex
 \end{center}
 \par
\vskip 2ex}

\newcommand{\preprintno}{UPR-783T}     

\def\abstract{\if@twocolumn
\section*{Abstract}
\else \absize
\begin{center}
{\bf Abstract\vspace{0pt}}
\end{center}
\setskip
\quotation
\fi}
\def\endabstract{\if@twocolumn\else\endquotation\fi}

\def\section{
\@startsection {section}{1}{\z@}{-3.5ex plus -1ex minus -.2ex}{2.3ex plus
.2ex}{\Large\bf}}

\def\tisize{\Large}     
\def\ausize{\large}     
\def\absize{\normalsize}     
\def\setskip{ \setlength{\baselineskip}{4ex} }     

\def\starttext{
\setlength{\baselineskip}{17pt}     
\pagenumbering{arabic}
}

\def\e{{\rm e}}
\def\beq{\begin{equation}}
\def\eeq{\end{equation}}
\def\ba{\begin{array}}
\def\bea{\begin{eqnarray}}
\def\ea{\end{array}}
\def\eea{\end{eqnarray}}

\def\Tr{ {\rm Tr}\, }
\def\Lk{\hbox{\sl Lk}}
\def\Wr{\hbox{\sl Wr}}
\def\Tw{\hbox{\sl Tw}}
\def\e{{\rm e}}
\def\kt{k_{\tiny B} T}
\def\eff{_{\rm eff}}
\def\p{_p}	
\def\init{_{\rm i}}	
\def\f{_{\rm f}}
\def\Hop{{\cal H}}
\def\Jop{{\cal J}}
\def\Aop{{\cal A}}\def\Aopp{{{\cal A}_+}}
\def\Bop{{\cal B}}\def\Bopp{{{\cal B}_+}}
\def\Nop{{\cal N}}
\def\inv{^{\raise.15ex\hbox{${\scriptscriptstyle -}$}\kern-.05em 1}}
\def\ie{{\it i.e.}}\def\eg{{\it e.g.}}
\def\sfbx{\hat e_{\rm x}}
\def\sfby{\hat e_{\rm y}}
\def\sfbz{\hat e_{\rm z}}
\def\bfbx{\hat E_1}
\def\bfby{\hat E_2}
\def\bfbz{\hat E_3}
\def\sfz{_{\rm z}}
\def\bfn{_1}	
\def\bfm{_2}	
\def\bft{_3}	
\def\ss#1{{\sf #1}}
\def\Mg{{\sf g}}
\def\ML{{\sf L}}

\def\bOmega{{\sf\Omega}}

\def\eqn#1{(\ref{#1})}  

\def\half{{\txt{1\over 2}}}

\begin{document}

\renewcommand{\@cite}[2]{$^{{\tiny #1}}$}

\setcounter{footnote}{1}
\title{Entropic Elasticity of Twist-Storing Polymers}

\author{J.~David Moroz\thanks{Address after
August 1, 1998: Department of Physics, University of California San
Diego 92093 USA}\hspace{5pt} and Philip Nelson\thanks{Corresponding author.}
\\[1ex]
\sl Department of Physics and Astronomy\\
\sl University of Pennsylvania\\
\sl Philadelphia, PA  19104\\[3ex]
}

\date{\vbox{\hbox{26 November 1997}\hbox{{\bf Revised 1 June 1998}}}}

\renewcommand{\preprintno}{~~}

\begin{titlepage}

\maketitle

\def\thepage {}        

\begin{abstract}
We investigate the statistical mechanics of a torsionally constrained
polymer.  The polymer is modeled as a fluctuating rod with bend
stiffness $A\kt$ and twist stiffness $C\kt$.  In such a model, thermal
bend fluctuations couple geometrically to an applied torque through
the relation $\Lk = \Tw + \Wr$.  We explore this coupling and find
agreement between the predictions of our model and recent experimental
results on single $\lambda$-DNA molecules.  This analysis affords an
experimental determination of the microscopic twist stiffness
(averaged over a helix repeat).  Quantitative agreement between theory
and experiment is obtained using $C=\Cvalue$ nm (\ie{} twist rigidity
$C\kt=$\Cvaluecgs ).  The theory further predicts a thermal reduction
of the effective twist rigidity induced by bend fluctuations. Finally,
we find a small reflection of molecular chirality in the experimental
data and interpret it in terms of a twist-stretch coupling of the DNA
duplex.

\end{abstract}

\end{titlepage}
\starttext

\catcode`\@=12
\def\cross{\!\times\!}
\def\mod{\hbox{\rm mod}~}
\def\pt{\partial}
\def\lt{\left}
\def\rt{\right}
\def\perpp{{\siriptscriptstyie\perp}}
\def\kb{k_{\scriptscriptstyle\rm B}} \def\tiny{\scriptscriptstyle\rm}
\def\der#1{{d{#1}(\ell)\over d\ell}}
\def\half{{1\over 2}}
\def\nml{\hbox{$\cal N$}}
\def\ham{\hbox{$\cal H$}}
\def\kbT{k_{\scriptscriptstyle\rm B}T}
\def\TT{\hbox{$\widetilde T$}}
\def\bo#1{{\cal O}(#1)}
\def\th{{\bf\hat t}}
\def\dnb{\delta\vec n}
\def\ofx{({\vec x})}
\def\ofxp{({\vec x}')}
\def\ofxt{}
\def\hef{$^4${\rm He}}
\def\het{$^3${\rm He}}
\def\lb{\hbox{$\bar\lambda$}}
\def\rsq{\hbox{$\overline{\langle\,\left( r(t) - r(0) \right)^2\,\rangle}$}}
\def\free{\hbox{$\cal F$}}
\def\bold#1{\setbox0=\hbox{$#1$}%
     \kern-.010em\copy0\kern-\wd0
     \kern.025em\copy0\kern-\wd0
     \kern-.020em\raise.0200em\box0 }

\def\grad{\bold{\nabla}}
\def\brac#1#2{\{{#1},{#2}\}}
\def\thth#1{\Theta\left[z-s_{#1}\right]\Theta\left[e_{#1} -z\right]}

\section{Introduction}

In this paper, we investigate the statistical mechanics of a polymer
chain with torsional rigidity.  We model the polymer as an elastic rod
subject to thermal fluctuations.  Each conformation of the chain is
statistically weighted according to the energy associated with bending
and twisting.  This is in contrast to conventional polymer models,
which account only for the energy cost of bending the polymer
backbone.\cite{DoiEdwards} This neglect of torsional energy is often
well justified, as many polymers are free to release twist by
swiveling about the single carbon bonds that constitute their
backbone.  Even for polymers that cannot swivel freely, the twist
usually amounts to an uncoupled Gaussian degree of freedom that can
simply be integrated away.  The situation is quite different, however,
in the presence of a torsional constraint.  In this case, the twist is
coupled to the conformation of the backbone and cannot be eliminated
so easily.  Such a situation can arise when the polymer is ligated
into a circle, or when its ends are clamped and a torque is applied at
one end. The concept of a torsional constraint can also be extended to
the {\it dynamics} of a polymer in a viscous fluid: here viscous
damping provides the necessary resistance to the
stress.\cite{kami98a,gold98a} Whatever the origin of the constraint,
it will result in a coupling between the twist and the bending modes
of the backbone.

The origin of this coupling lies in White's theorem:
$\Lk=\Tw+\Wr$.\cite{Calugareanu,White,Fuller} This formula relates a
global topological invariant of any pair of closed curves (the Linking
number, $\Lk$), to the sum of a local strain field (the Twist, $\Tw$)
and a global configurational integral (the Writhe, $\Wr$).  If the
linking number is fixed, the polymer will be forced to distribute the
invariant $\Lk$ between the degrees of freedom associated with $\Tw$
and $\Wr$.  From a statistical mechanics point of view, the set of
complexions available to the system is then restricted.  The elastic
energy of each allowed complexion reflects the sum of a twisting
energy and a bending energy associated with the Writhe of the
backbone.  Of course we do not need to consider fixed linking number
for torsional rigidity to be important: a chemical potential for $\Lk$
in the form of an applied torque also couples the bend fluctuations to
the twist.

Perhaps the most important examples of twist-storing polymers are
biopolymers, especially DNA.  Unlike many of its hydrocarbon-chain
cousins, the monomers of DNA are joined by multiple covalent bonds;
additional specific pairing interactions between bases prevent
slippage between the strands.  This multiply-bonded structure inhibits
the unwinding of the DNA helix to release a torsional stress; instead,
there is an elastic energy cost associated with the deformation.

Recently it has become possible to perform experiments on single
molecules of DNA.  In a classic experiment, Smith {\it et
al.}\cite{Smith} anchored one end of a DNA duplex to a solid substrate
while the other end was attached to a magnetic bead.  The
conformations of the polymer could then be probed by considering the
end-to-end extension of the chain as a function of the magnetic force
applied to the bead.  These experiments, and others which stretch DNA
molecules using electric fields,\cite{Schurr1990} hydrodynamic
flows,\cite{Perkins1990} or optical tweezers\cite{Block} were soon
analyzed using the ``worm-like chain'' (WLC) model.\cite{DoiEdwards}
Working within this framework, Bustamante, Marko and
Siggia\cite{bust94a,MarkoSiggiaMacro} and Vologodskii\cite{volo94a}
were able to reproduce the experimental force-extension curves for DNA
over a wide range of forces (from $10^{-2}$ pN to $10$ pN) with just
one fitting parameter, the DNA bend persistence length.

Since the original DNA stretching experiments, significant
improvements have been made.  In particular, a series of elegant
experiments\cite{Strick,StrickSupercoil,Allemand98} has succeeded in
{\it torsionally} constraining the DNA using swivel-free attachments
at both ends.  As a result, one can now directly explore the interplay
between DNA's internal resistance to twisting and the conformations of
its backbone.

In this paper, we will explain some of these new results analytically
in terms of a theory of twist-storing polymers.  Our final formula,
given in 
\eqn{finalz} below, quantitatively fits the experimental data of
Strick {\it et al.}\cite{Strick} and of Allemand and
Croquette\cite{Allemand98} with just two important fit parameters: the
bend stiffness $A$ and twist stiffness $C$ (a more precise statement
appears below). Our analytical approach rests upon linear elasticity
and perturbation theory about a straight rod. Thus we do not address
the remarkable structural transitions induced in DNA by torsional
stress,\cite{Strick,StrickSupercoil} nor will we systematically study
the plectonemic transition or other phenomena involving
self-avoidance.  Marko and Siggia have previously studied the effects
of thermal fluctuations on plectonemic DNA;\cite{MarkoSiggiaPRE} we
have chosen instead to work in a regime not afflicted by this
theoretical difficulty.  We will show that our analysis is justified
in a well-defined region of parameter space where many experimental
data points are available (solid symbols in Figure~\ref{figure1}), and
from the data deduce the fundamental elastic parameters of DNA.

The main points of our results were announced
previously.\cite{PhilDIMACS,JDMPNPNAS} Some of the steps were
independently derived by Bouchiat and M\'ezard\cite{BouchiatMezard} in
a different analysis of the same experiments.  The present paper gives
some new analytical results, particularly in section~\ref{nonpert},
and applies the analysis to some new experimental data (see
Figure~\ref{figure1}).

In addition to these analytical results, Vologodskii and Marko, and
Bouchiat and M\'ezard, have recently performed Monte Carlo
simulations\cite{MarkoVolo,BouchiatMezard} to study the conformations
of DNA under applied tensions and torques appropriate to those in the
experiments studied here.  Marko has also studied the related problem
of torsional constraints on the overstretching
transition.\cite{MarkoTwistStretch,MarkoHighTension}

Apart from quantitatively reproducing the experimental extension
curves with just a few fit parameters, our theory also predicts a
reduction of the effective twist rigidity of a polymer caused by
conformational fluctuations.  We give the form of a new effective
twist rigidity $C\eff\kt$, which is smaller than the microscopic
rigidity $C\kt$.  This effect, anticipated some time ago by Shimada
and Yamakawa\cite{ShimYama} has a simple explanation: part of the
excess Link imposed on a solid rod can be moved into the bend
deformations of its backbone through the coupling associated with the
$\Lk$ constraint.  Our simple formula (\eqn{Ceff} below) makes this
intuition precise for the case of a highly stretched rod.

It may at first seem that all the relevant physics could be found in
the classical works of the nineteenth century,\cite{Love} but actually
one can see at once that classical beam theory is qualitatively at
odds with the experimental data of Figure~\ref{figure1}: it says that
a rod under tension will simply twist in response to an applied torque
$\tau$ as long as $\tau$ is small enough.  Only when the torque
exceeds a critical value will the rod buckle into a helical
configuration, thus shortening the end-to-end extension.  Unlike its
macroscopic counterpart, however, a microscopic rod is continuously
buffeted by thermal fluctuations.  Because the rod is never straight,
its average shape will respond as soon as any torsional stress is
applied; there is no threshold, as seen in Figure~\ref{figure1}. In
sections~\ref{physicalpicture}--\ref{maincalc} we will create a simple
mathematical model embodying this observation and use it to explain
the data.

\section{Experiment\label{pnSexperiment}}

The statistical mechanical problem of a twist-storing polymer subject
to a $\Lk$ constraint is realized in the experiments of Strick {\sl et
al.}\cite{Strick,StrickSupercoil} and Allemand and
Croquette.\cite{Allemand98} In these experiments, a segment of
double-stranded $\lambda$-DNA of length $L\approx 15.6\,\mu$m is held
at both ends: one end is fixed to a glass plate while the other is
attached to a magnetic bead.  Both ends are bound in such a way as to
prevent swiveling of the polymer about the point of attachment.  By
rotating the magnetic bead in an applied magnetic field, the
experimenters are then able to adjust the excess linking number to any
desired, fixed value.

While the direction of the applied field fixes the linking number, a
gradient in the same field allows the DNA molecules to be put under
tension.  The experiment is therefore able to study the statistical
mechanics of the biopolymer in the fixed tension $f$ and linking
number $\Lk$ ensemble.  The measured response is then the end-to-end
extension $z(f,\Lk)$ of the chain as a function of the applied stress.
In contrast, traditional ligation experiments control only $L$ and
$\Lk$, and $\Lk/L$ can take on only rather widely-spaced discrete
values. Moreover, the measured quantity is gel mobility, whose
relation to backbone conformation is not simple.

Some of the experimental results for forces greater than 0.1~pN are
shown in Figure~\ref{figure1}. In the figure, the solid lines are our
theoretical fit to the solid points.  These curves were produced by
fitting four parameters: the microscopic persistence lengths $A,C$ and
twist-stretch coupling $D$ (all averaged over a helical repeat), as
well as the arclength of the polymer $L$.  The bend persistence length
$A$ has been determined in a number of earlier
experiments,\cite{Smith,cluzel,Block} while $L$ can be determined from
only the data points with zero excess Link.  The fitted values of $A$
and $L$ therefore serve mainly as a check of the theory.  In our fit
we used \Nvalue\ different points, only some of which are depicted as
the solid symbols in Figure~\ref{figure1}.  The figure also shows open
symbols.  These points correspond to ($f$,$\Lk$) pairs that lie
outside the region where our model, which has no explicit
self-avoidance, is valid.  Due to this neglect of self-avoidance, our
{\sl phantom chain} model will have a mathematical pathology
associated with configurations that include self-crossings.  To deal
with these difficulties, we will simply require that the chain be
pulled hard enough that such configurations become statistically
negligible.  As we will see, ``pulling hard enough'' corresponds to a
restriction on the applied stretching force $f$ and the applied torque
$\tau$ (see appendix~B).  Apart from the restrictions of the {\sl
phantom chain} model, there were also omissions of data points for
physical reasons.  For example, at large applied tensions and torques,
the DNA molecule undergoes structural transformations.  In
section~\ref{fitresults}, we will discuss our data selection criteria
and the fitting procedure more fully.

\section{Physical Model \label{physicalpicture}}

Throughout most of this paper we will model DNA as a fluctuating
elastic rod of uniform circular cross-section and fixed contour length
$L$.  This idealization neglects DNA's helical nature: in particular,
the length scale associated with the helical pitch of the molecule
($2\pi/\omega_0=3.6\,$nm) does not enter as a parameter.  The concept
of {\sl fractional overtwist} ($\sigma=2\pi\Delta\Lk/L\omega_0$) is
therefore meaningless.  Nevertheless, we will retain the traditional
notation to provide a connection to the published experimental data,
expressing our results in terms of $\sigma$ and noting that $\sigma$
and $\omega_0$ enter only in the combination $\sigma\omega_0$.  In the
main text we will show that our achiral, isotropic elastic rod model
captures the main features of Figure~\ref{figure1}.  At the end of our
calculation, in \eqn{finalz}, we will also allow for intrinsic
stretching and a possible asymmetry between positive and negative
$\sigma$, a chiral effect associated with the twist-stretch coupling
of a helical rod.

In appendix~A we will introduce helical pitch effects and show that at
modest stretching tension they can be summarized in an effective,
``coarse-grained'' energy (see \eqn{bendtwist} below). They also lead
to a new phenomenon, {\sl chiral entropic elasticity}, via the
twist-bend coupling of DNA.\cite{MarkoSiggiaBendTwist} This effect is
potentially another source of asymmetry between over- and
undertwisting, but the available data do not at present give detailed
information about the asymmetry, and so we omit this complication from
the main text.

Accordingly we define an elastic energy functional which describes the
bending and twisting of an isotropic elastic rod of fixed arclength
$L$:\cite{Landau}
\beq
{{E_{\rm bend}}\over\kt}={A\over 2}\int_0^{ L}\
(\dd\hat t/\dd s)^2
\dd s,\qquad{\rm and}\qquad
{{E_{\rm twist}}\over\kt}={C\over 2}\int_0^{ L}\ {\Omega\bft}^2 \ \dd s .
\label{bendtwist}
\eeq
In these formulas $\hat t(s)$ is the tangent to the rod backbone at
the point with arclength $s$ from the end. We imagine inscribing
permanently a ``material frame'' embedded in the rod; then
$\Omega\bft$ is the rate of rotation of this frame about $\hat t$ (see
\eqn{pnomdef} below; our notation mainly follows that of Marko and
Siggia\cite{MarkoSiggiaPRE}). We are free to choose a convenient
material frame; we choose one which coincides with the fixed lab frame
when the molecule is unstressed. (In keeping with the remarks above,
there is no reason to choose a material frame initially rotating
relative to the lab at $\omega_0$.)  $A$ and $C$ are the bend and
twist ``persistence lengths,'' which are given by the respective
elastic constants divided by $\kt$. These parameters are understood to
be averaged (or ``coarse-grained'') over the scale of a helical
repeat. In appendix~A we find the relation between them and a more
elaborate elasticity theory incorporating the intrinsic helicity of
the DNA duplex.

Equations \eqn{bendtwist} are mathematically identical to the {\it
kinetic} energy of a {\it symmetric spinning top} with arclength $s$
playing the role of time.  Hence there is a direct analogy between the
{dynamical equations of motion for a top} and the equations describing
the {equilibrium for an elastic rod}, an observation due to
Kirchoff.\cite{Kirchoff} The main technical point of our analysis is
the extension of Kirchoff's observation to a mathematical
correspondence between the {\it thermal fluctuations} of an elastic
rod and the {\it quantum mechanics} of a spinning
top.\cite{BouchiatMezard,JDMPNPNAS,PhilDIMACS}

The bend persistence length $A$ which appears in \eqn{bendtwist} is a
well-known parameter that has been measured in several experiments.
Among other things, this parameter is known to depend on the salt
concentration of the surrounding fluid.\cite{CMeasure1} Wang {\it et
al.} have measured $A=47\,$nm for DNA in buffer conditions similar to
those in the experiments studied here.\cite{Block}

The value of the twist persistence length $C$ has not been determined
as directly as $A$.  Cyclization kinetics
studies,\cite{Shore1983b,Taylor1990,ShimYama,Cmeasure3} topoisomer
distribution analyses\cite{Shore1983a,Horowitz1984} and fluorescence
polarization anisotropy (FPA)
experiments\cite{HeathJMB,Fujimoto1990,Schurr1992} have provided
measurements of this parameter, but these determinations are somewhat
indirect and the results have been difficult to reconcile with each
other.\cite{CMeasure1,CMeasure2} In particular, results obtained from
straight and circular DNA's using a single technique (FPA) yield
different values of the twist rigidity: $C\approx 50$ nm for linear
DNA's and $C\approx 85$ nm for circular DNA's.\cite{HeathJMB} This
discrepancy may be a consequence of the thermal softening of the
torsional rigidity predicted by our theory (see \eqn{Ceff}).  The main
goal of the present paper is to interpret the single DNA molecule data
in Figure~\ref{figure1} in terms of a theory we call ``torsional
directed walks'', thereby permitting a new measurement of $C$.  Like
the bending rigidity $A$, $C$ may be expected to depend on the buffer
solution; the dependence of $C$ should however be much weaker than $A$
since twisting does not modify the spatial distribution between
charges on the backbone to the same degree as bending.

The rod is subject to a stretching force $f$ and a torsional
constraint.  It will prove simplest to impose the torsional constraint
through a fixed applied torque $\tau$ rather than directly through a
fixed linking number.  Since the molecules we will study are many
times longer than $A$ or $C$, we are in the thermodynamic limit, and
so we expect the two ensembles to give the same physical results.

The two stresses on the polymer require the introduction of
two more terms in the polymer's energy functional:
\beq
{{E_{\rm tension}}\over\kt}=- \tilde{f}\cdot z
=-\tilde{f}\int_0^L\hat t\cdot\sfbz\dd s
,\qquad{\rm and}\qquad
{{E_{\rm torque}}\over\kt}=- 2 \pi \tilde{\tau}\cdot  \Lk .
\label{lagrangeterms}
\eeq
Here $z$ is the end-to-end extension of the polymer.  The tension and
torque have been expressed in terms of the thermal energy:
\beq
\tilde{f}\equiv f/k_{\tiny B} T,\quad{\rm and}\quad
\tilde{\tau}\equiv \tau/k_{\tiny B} T.\eeq
In \eqn{lagrangeterms} and throughout this paper, $\Lk$ denotes the
{\it excess} Link, consistent with the remarks at the beginning of
this section; thus $\Lk=0$ for the unstressed rod. In general, $\Lk$
is defined only for {\it closed} loops. If we have an {\it open} chain
with both ends held at fixed orientations, as in the experiments under
study, then we can draw a fixed, imaginary return path completing our
chain to a closed loop and let $\Lk$ denote the Link of this closed
loop. Choosing the return path so that $\Lk=0$ when the rod is
straight and unstressed then gives in general $\Lk=\Tw+\Wr$ where the
terms on the right refer only to the open, physical rod.

Before we include $E_{\rm torque}$ in our energy functional, the Link
must be more explicitly expressed.  To get a useful expression, we
first note that the Twist is defined as
\beq \Tw={1\over2\pi}\int_0^L\Omega\bft\dd s .
\eeql{pntwist}%
The Writhe involves only the space curve $\vec{r}(s)$ swept out by the
rod's centerline.  In general, this number is given by a complicated,
non-local formula\cite{Calugareanu,FullerEarly,White} involving a
double integral around the closed curve:
\beq
\Wr = {1\over {4\pi}}\oint \dd s \oint \dd s' \left(
{{\dd\vec{r}(s)}\over{\dd s}} \cross {{\dd\vec{r}(s')}\over{\dd
s}}\right)
\cdot {{\vec{r}(s)-\vec{r}(s')}\over{\vert\vec{r}(s)-\vec{r}(s')\vert^3}}.
\label{doubleint}
\eeq
However, a result due to Fuller allows us to rewrite this quantity as
a single integral over a local Writhe density.  This simplification is
made possible by noting that for small variations about some reference
curve $\vec{r}_0(s)$, the integrand in \eqn{doubleint} becomes a total
derivative.  Performing one of the integrals then yields a single
integral over a local quantity.\cite{Fuller} Specializing to the case
where the reference curve is just the $\sfbz$-axis then
gives\cite{RudnickLinear}
\beq
\Wr={1\over{2\pi}}\int {\hat{t} \cross \sfbz \cdot {\dd\hat t/
\, {\dd s}}\over{1+\hat{t}\cdot\sfbz}} \dd s.
\label{pnwrithe}\eeq
Fuller's result holds as long as there is a continuous set of
non-self-intersecting curves interpolating between the reference curve
and the curve in question, {such that the denominator in
\eqn{pnwrithe} never vanishes.}  We can now combine the terms to get
the full energy functional for our model of DNA:
\beq
{E\over\kt}={E_{\rm bend}\over\kt} + {E_{\rm twist}\over\kt}
-\tilde{f} \cdot z - 2\pi \tilde{\tau}\cdot \Lk .
\label{torsionenergy}
\eeq

Formul\ae{}
(\ref{bendtwist},\ref{lagrangeterms},\ref{pntwist},\ref{pnwrithe},
\ref{torsionenergy}) define
the elastic model we will use through the end of
section~\ref{nonpert}. Later, in section~\ref{fitresults} and
appendix~A we will consider various elaborations of the model and
determine that they are relatively unimportant in capturing the main
features of the data in Figure~\ref{figure1}.

As noted in the introduction, we expect that thermal fluctuations will
have an important effect on the rod's twist degree of freedom.  A
macroscopic elastic rod under tension will sustain a finite amount of
applied torsional stress without buckling.  Once a threshold is
reached, however, the stress can be partially relaxed by bending the
backbone.  Linear stability analysis of the energy \eqn{torsionenergy}
shows that this threshold is given\cite{Love} by $\tilde{\tau}_{\rm
crit}=2\sqrt{A\tilde{f}}$.  Unlike its macroscopic counterpart,
however, a microscopic rod is subject to thermal fluctuations.  These
fluctuations prevent the rod from ever being straight; as we show
below, even infinitesimal torsional stresses will then affect the bend
fluctuations.  Even though there is no chiral energy term, individual
fluctuations will not be inversion symmetric.  An applied torsion will
push the fluctuations with the corresponding helical sense closer to
instability, while suppressing those of the opposite helical sense.
The end result will be a coupling between the applied torsion and the
mean end-to-end extension of the rod proportional to $\tau^2$ (terms
linear in $\tau$ must drop out since the model does not break
inversion symmetry).

Later we will consider the effects of molecular chirality: \eg{} in
section~\ref{fitresults}, we will include a twist-stretch coupling
term $D$.\cite{MarkoTwistStretch,KLNO1,KLNO2} It will turn out that
the effect of this coupling on the experiment we study is small: this
is already apparent in Figure~\ref{figure1} where the data points are
nearly symmetric about $\sigma=0$.  Nevertheless, by including the
twist-stretch coupling, we will be able to determine the parameter $D$
roughly.

Another way that chirality enters a physical model of DNA is through
an anisotropic bending term.  Any transverse slice through the
molecule is easier to bend in one direction than in another.
Microscopically, this anisotropy has its origin in the shape of the
base pair plates that make up the rungs on the DNA ladder.  Since
these plates are longer in one direction than the other, bending about
the short axis (``tilt'') is more difficult than bending about the
long axis (``roll'').\cite{ulya84a,srin87a,sara89a} In appendix~A we
consider such an anisotropy, as well as the related twist-bend
coupling,\cite{MarkoSiggiaBendTwist} finding that these effects can be
summarized to good accuracy in an effective coarse-grained model of
the form \eqn{bendtwist}.  This conclusion could have been anticipated
since the important fluctuations are on length scales around
$2\pi\sqrt{A/\tilde{f}}$, and for the forces below $8$~pN that we
consider, this averages over at least several helical repeats.  We
conclude that the treatment of DNA as an achiral rod of elastic
material is sufficient to understand how its extension changes under
applied tension and torque.

At this point it may be noted that unstressed natural DNA is {\it not}
a perfect helix; its axial symmetry is already broken, even in the
absence of thermal fluctuations.  In particular, it is well known that
the unstressed, zero temperature structure of DNA is sequence
dependent.\cite{trif87a,sche95a} The effect of this quenched disorder
has been studied recently by Bensimon, Dohmi, and
M\'ezard\cite{BensimonDisorder} and by one of us.\cite{NelsonDisorder}
For simple models of weak disorder, the main effect is simply to
renormalize the bend persistence length $A$.  In the present paper, we
neglect explicit inclusion of the quenched disorder associated with
sequence-dependent effects.  Thus our bend rigidity $A$ is the
effective value including disorder.

Even though the bend and twist rigidities represent averages over a
helix repeat, they are still microscopic parameters and therefore
reflect only the short-scale behavior.  As we go to longer length
scales, we expect the effective bend and twist rigidities to be
modified by the geometric coupling implicit in White's formula.  In
particular, we will find that the effective twist rigidity is reduced
for small applied tensions:
\beq
C\eff = C \left( 1+ {C\over{4A\sqrt{A\tilde{f}}}}\right)^{-1}.
\label{Ceff}
\eeq
The dependence of $C\eff$ on length scale enters through $\tilde{f}$:
as mentioned above, $\sqrt{A/\tilde{f}}$ sets the scale of the most
important fluctuations in the problem.  At small tensions, or
equivalently at long length scales, $C$ is effectively reduced.
Equation \eqn{Ceff} describes this ``softening'' of the twist
rigidity.  The reducing factor is explicitly dependent on $\kt$,
indicating that this is a thermal effect.

\section{Group Language \label{grouplanguage}}

In the next section we will consider the thermodynamic complexions
available to a torsionally constrained polymer.  To prepare for the
task, we must first define convenient variables for evaluating the
energy functional of the last section on the group of rotations,
$SO(3)$.  The bending and twisting deformations that appear
in \eqn{bendtwist} as well as the Lagrange multiplier terms for
extension and Link which appear in \eqn{lagrangeterms} will need to be
expressed in terms of these variables.

We will use two reference frames related by an element of the rotation
group.  The first of these frames is ``space-fixed''; we will take as
its basis the orthonormal triad $\{\hat{e}_{i}\}$, with $i=$x,y, or z.
A rotation $\Mg(s)$ relates this frame to the ``body-fixed'' (or
``material'') frame $\{\hat{E}_{\alpha}(s)\}$ with $\alpha=1,2$, or 3,
where $s$ denotes a point on the rod backbone.  As mentioned earlier,
we will take $\hat E_3(s)=\hat t(s)$ to be the tangent to the rod's
centerline, and the remaining two vectors to be constant directions
when the rod is straight and unstressed.  The local orientation of the
polymer is then given by the $3 \cross 3$ orthogonal matrix
$\Mg_{\alpha i}(s)=\hat{E}_\alpha(s)\cdot\hat{e}_i$. The matrix $\Mg$
contains only three independent entries. We will sometimes find it
convenient to represent it in a nonredundant way using Euler angles:
\bea
\Mg(s)&=&\e^{-\ML_3 \psi(s)} \e^{-\ML_1 \theta(s)} \e^{-\ML_3 \phi(s)} .
\label{pneuler}
\eea
Thus for example $\hat{t}(s)\cdot\sfbz=\Mg_{3{\rm
z}}(s)=\cos\theta(s)$.

The generators of infinitesimal rotations are then matrix operators
acting on $\Mg$.  When these operators act from the left they are
called ``body-fixed rotations''; when they act from the right they are
called ``space-fixed rotations''. In either case a convenient basis
for the generators is
\bea
\ML_1=\left(\ba{ccc} 0&0&0\\ 0&0&1 \\ 0&-1&0 \ea\right),\qquad
\ML_2=\left(\ba{ccc}
0&0&-1\\0&0&0\\1&0&0 \ea\right),\qquad{\rm and}\qquad
\ML_3=\left(\ba{ccc} 0&1&0\\-1&0&0\\0&0&0 \ea\right).
\eea
We can then describe the rotation of the material frame as we walk
along the rod backbone as an infinitesimal body-fixed rotation
$\bOmega$ or as a space-fixed rotation $\hat{\bOmega}$, where
\beq
\bOmega=\dot\Mg\Mg\inv\quad {\rm and}\quad \hat{\bOmega}=\Mg\inv\dot\Mg
.\label{pnomdef}\eeq Here and elsewhere, a dot signifies $\dd/\dd s$.
We will also write the projections of the rotation rates onto the
generators as
\beq
\Omega_\alpha\equiv(\bOmega,\ML_\alpha)\equiv
-\half\Tr[\bOmega\ML_\alpha]
\label{pnomcomp}\eeq
and similarly for $\hat{\Omega}_i$.

With these definitions we can cast the formulas of the previous
section into more useful forms. We first compute that $(\dd\hat t/\dd
s)^2={\Omega\bfn}^2+{\Omega\bfm}^2$ and substitute into
\eqn{bendtwist}. Next, a simple calculation gives
\beq
\Omega\bft=-(\dot \psi+\cos\theta\dot\phi)
\quad{\rm and}\quad
\hat\Omega\sfz=-(\dot\phi+\cos\theta\dot\psi)
.\label{pnthrees}\eeq Next, note that $\hat t=\bfbz=\Mg_{3i}\hat e_i=
\sin\theta(\sin\phi\sfbx+\cos\phi\sfby)+\cos\theta\sfbz$.
Explicit evaluation of the local Writhe density \eqn{pnwrithe} then
gives with (\ref{pntwist},\ref{pnthrees}) that
\beq
\Lk=-{1\over2\pi}\int(\dot\psi+\dot\phi)\,\dd s=
{1\over2\pi}\int{\Omega\bft+\hat\Omega\sfz
\over 1+\cos\theta}\,\dd s .
\label{pnLkOm}\eeq
With this last expression, the energy functional \eqn{torsionenergy}
is explicitly given in terms of an element of the rotation group and
its derivatives, as expressed by the angular frequencies
$\Omega_=\alpha$ and $\Omega_{\rm z}$.

We close this section with a mathematical fine point, which will not
affect our calculation. Strictly speaking, our configuration space is
only locally the group manifold $SO(3)$. We will exclude the points
$\theta=\pi$ where \eqn{pnLkOm} is singular.  Moreover, we need to
``unwrap'' the remaining space. The physical origin of this step is
simply the fact that rotating the rod by $2\pi$ does not return it to
an equivalent state, but rather introduces an extra unit of
Link. Mathematically we simply remember that $\phi+\psi$ is not to be
identified modulo $2\pi$ (see
\eqn{pnLkOm}), though $\phi-\psi$ is.

\section{Calculation \label{maincalc}}

\subsection{The Path Integral \label{calculation}}

We wish to compute the average extension $\langle z \rangle$ and
relative excess Link $\langle \Lk \rangle$ for a twist-storing polymer
subject to a given tension and torque.  To find these properties, we
must first compute the partition function.  At each point along the
arclength of the polymer, the local orientation will be given by some
rotation $\Mg$.  To calculate the weight of any configuration entering
into the partition function, we simply apply the appropriate Boltzmann
factor.  In the last section we described how the terms of the energy
functional appearing in this factor can be written in terms of
rotations.  Using these expressions, it is now possible to write down
a path integral on the group space:
\beq
{\cal Z}=\int [\dd \Mg(s)]\ \exp\left(-{1\over\kt}\left(E_{\rm
bend}+E_{\rm twist}\right)+ 2\pi \tilde{\tau}\cdot \Lk+\tilde{f} \cdot
z\right) .
\eeql{partition}
This partition function gives us the quantities of interest, namely
the average chain extension $\langle z\rangle$ and the average excess
Link resulting from an applied tension and torque:
\beq
\langle z\rangle= \left.
{\partial\over{\partial\tilde{f}}}\right\vert_{\rm \tilde{\tau}}
\ln{\cal Z},\qquad
\langle\Lk\rangle=\left.
{1\over{2\pi}}{\partial\over{\partial\tilde{\tau}}}\right\vert_{
\tilde{f}} \ln{\cal Z}.
\eeql{extension.link}
A direct evaluation of the partition sum in \eqn{partition} is
difficult; fortunately, such an evaluation proves to be unnecessary.
In this paper we instead extend a standard polymer physics
trick.\cite{DoiEdwards,MarkoSiggiaMacro} It turns out that the
partition sum is closely related to the ``propagator'' for the
probability distribution for the polymer's orientation $\Mg$.  We
define the unnormalized propagator by
\beq
\Psi(\Mg\f ,s\f;\Mg\init,s\init)=\int_{{\Mg(s\f)=\Mg\f} \atop {\Mg(s\init)=\Mg\init}}
[\dd \Mg(s)]\ \exp\left(-{{E[\Mg(s)]}\over\kt}\right) .
\label{pathintegral}
\eeq
The probability $P_{s}(\Mg)$ for the polymer to have orientation $\Mg$
at position $s$ is then given by a multiplicative constant times $\int
\dd\Mg\init\ \Psi(\Mg,s;\Mg\init,0) P_{s=0}(\Mg\init)$.  More interestingly from our
perspective, for a long chain $\log\Psi(\Mg,L;\Mg\init,0)$ becomes
independent of $\Mg$ and $\Mg\init$.  In fact the propagator is then
just a constant times the partition function ${\cal Z}$.  The utility
of studying the seemingly complicated $\Psi$ instead of ${\cal Z}$
comes from the realization that $\Psi$ obeys a simple differential
equation.  We will derive this equation in section~\ref{SE}.  Its
solution for large $L$ is dominated by a single eigenfunction of the
differential operator.  Armed with this knowledge, we will compute in
section~\ref{Quantities} quantities such as the average extension
$\langle z\rangle$ and linking number $\langle\Lk\rangle$ by
substituting $\Psi$ for ${\cal Z}$ in the thermodynamic relations
\eqn{extension.link}.

\subsection{The Schr\"odinger-Like Equation\label{SE}}
The next step, then, is to determine the differential equation obeyed
by $\Psi(\Mg,s;\Mg\init,0)$ as a function of $s$.  To do
this,\cite{FeynmanHibbs} consider the evolution over a short backbone
segment of length $\epsilon$:
\beq
\Psi(\Mg\f,s\f+\epsilon;\Mg\init,0)=\int\dd \Mg_1 \left[{1\over N} \int_{{
\ss h(s\f+\epsilon)=\Mg\f}\atop{\ss h(s\f)=\Mg_1}} [\dd \ss h(s)]\
\exp\left(-{{\delta
E[\ss h(s)]}\over\kt}\right)\right] \Psi(\Mg_1,s\f;\Mg\init,0) .
\eeql{path}%
Here $\delta E[\ss h(s)]$ is the elastic energy of the short segment
of rod from $s\f$ to $s\f + \epsilon$.  We introduced a normalizing
factor $N$ to get a continuum limit: as long as this factor does not
depend on $\tilde{f}$ or $\tilde{\tau}$ it will not enter the
quantities of interest (see \eqn{extension.link}).  In this subsection
we will compute the functional integral in \eqn{path}, retaining terms
up to first order in $\epsilon$, and hence compute $\dd\Psi/\dd s\f$.

As $\epsilon \rightarrow 0$, we will see that only matrices $\Mg_1$
close to $\Mg\f $ produce appreciable contributions to the path
integral.  It is therefore possible to write $\Mg _1$ uniquely in the
form $\Mg _1=\exp(-T_\alpha
\ML_\alpha)\Mg\f$. Moreover, over the short segment under consideration we may
take $\ss h(s)$ to interpolate between $\Mg\f$ and $\Mg_1$ in the
simplest way:
\beq
\ss h(s)=\exp\left( {{s-s\f-\epsilon}\over{\epsilon}} T_\alpha \ML_\alpha
\right) \Mg \f.
\eeql{pnleft}%
The functional integral then reduces to an ordinary integral over
$\vec{T}$:
\beq
\int\dd \Mg_1\ \int_{{\ss h(s\f+\epsilon)=\Mg\f}
\atop{\ss h(s\f)=\Mg_1}} [\dd \ss h(s)] \rightarrow
\int\exp\left(-{{\vert\vec{T}\vert^2}\over 12}\right) \dd^3\vec{T} .
\label{pntrunc}\eeq
We have suppressed an overall constant, absorbing it into $N$ in
\eqn{path}.
The exponential factor on the right side gives the invariant volume
element of group space\cite{Tung} near the point $\Mg \f$.  In the
end, this factor will not modify the differential equation that we
develop, but it is included here for completeness.

The energy functional $\delta E[\ss h(s)]$ can now be evaluated on the
arclength slice of length $\epsilon$.  With the useful abbreviation
\beq
M_\alpha(\Mg\f)\equiv(\Mg\f\ML_3\Mg\f\inv,\ML_\alpha)=(\sin\theta\sin\psi,
-\sin\theta\cos\psi,\cos\theta),
\eeql{pnMdef}%
we get that $\Omega_\alpha=-T_\alpha/\epsilon$ and
$\hat\Omega\sfz=-\vec{M}\cdot\vec{T}/\epsilon$ which are constants
(independent of $s$) over the short segment. Thus
\beq
{{\delta E[\ss h(s)]}\over\kt}={A\over2\epsilon}({T_1}^2+{T_2}^2)
+{C\over2\epsilon}{T_3}^2+\tilde\tau\left(T_3 +{T_1M_1+T_2M_2\over
1+M_3}\right)-\epsilon\tilde{f}\cos\theta .
\eeql{pnenergyT}%
The factor $\e^{-\delta E/\kt}$ weights each path from $\Mg _1 =\exp
(- \vec{T}\cdot\vec{L}\,) \Mg\f$ to $\Mg\f $; as $\epsilon\rightarrow
0$ it indeed kills all those $\Mg _1$ which wander too far from $\Mg\f
$, {\it i.e.} all deformations where
$T_\alpha\grsim\sqrt{\epsilon/A}$.

We also need to express $\Psi(\Mg_1)$ in terms of $\vec{T}$. Here and
below we abbreviate $\Psi(\Mg\f,s\f,\Mg\init,0)$ by $\Psi(\Mg\f)$.
Define the left-acting (body-fixed) derivatives $\Jop _\alpha$ via
\beq
\Jop _\alpha\Psi(\Mg)\equiv[\ML_\alpha\Mg]_{\beta i}
\left.{\partial\Psi\over\partial\Mg_{\beta i}}\right|_\Mg,
\label{pndefJ}\eeq
and similarly the right-acting (space-fixed) derivatives $\hat{\Jop
}_i$.  Then $\Psi(\Mg_1)=\e^{-T_\alpha \Jop_\alpha}\Psi(\Mg\f)$ or
\beq
\Psi(\Mg_1)=
\Psi-\vec{T}\cdot\vec{\Jop }\,\Psi+\half T_\alpha T_\beta \Jop _\alpha \Jop
_\beta\,
\Psi+\cdots
,\label{pnPsi}\eeq where we abbreviated still further by omitting the
basepoint $\Mg\f$ on the right-hand side.

We can now combine
(\ref{path},\ref{pntrunc},\ref{pnenergyT},\ref{pnPsi}) and perform the
Gaussian integral $\dd^3\vec{T}$. First complete the square, defining
$\bar{T}_3=\sqrt{C/2\epsilon}\bigl(T_3+\epsilon\tilde\tau/C\bigr)$ and
$\bar{T}_\alpha=\sqrt{A/2\epsilon}\left(T_\alpha+
\epsilon\tilde\tau M_\alpha/A(1+\cos\theta)\right)$, $\alpha=1,2$.
Choose the normalization $N$ so that the limit $\epsilon\to0$
reproduces $\Psi$.  Collecting all order-$\epsilon$ terms and using
${M_1}^2 +{M_2}^2=\sin^2\theta$ then gives
\bea
\dot{\Psi}&=&\Biggl\{
{\tilde\tau^2\over2}\left({1\over C}+{1\over A}{1-\cos\theta\over1+
\cos\theta}\right) +\tilde{f}\cos\theta+{\tilde\tau\over
A(1+\cos\theta)}\bigl( M_1\Jop _1+M_2\Jop _2+M_3\Jop _3+\Jop _3\bigr)
\nonumber\\
&&+\tilde\tau \Jop _3\left({1\over C}-{1 \over A}\right)
+\half\left({1\over A}({\Jop _1}^2+{\Jop _2}^2)+{1\over C}{\Jop
_3}^2\right)
\Biggr\}\Psi.
\label{pnSE1}\eea
Further consolidation then gives $\dot{\Psi}=-\left(\Hop +{\cal
E}_0\right)\Psi$, where
\beq
{\cal E}_0\equiv -\left(\tilde f+{\tilde\tau^2\over 2C}\right)
\eeq
and the differential operator $\Hop$ is defined by
\bea
\Hop &=&{K\over A}\Biggl[
-{1\over 2K}\vec{\Jop }^{\,2}
+K(1-\cos\theta)-{\tilde\tau^2\over4K}{(1-\cos\theta)^2\over1+\cos\theta}
-{1\over 2K}\Bigl({A\over C}-1 \Bigr){\Jop _3}^2
\nonumber \\
&& -{\tilde\tau\over K}\left[\Bigl({A\over C}-\half\Bigr)\Jop \bft
+\half\hat \Jop \sfz\right] -{\tilde\tau\over
4K}{1-\cos\theta\over1+\cos\theta}(\Jop \bft+\hat{\Jop }\sfz)
\Biggr]
.\label{pnHdef}\eea We have arranged the terms in \eqn{pnHdef} to
facilitate a systematic expansion in powers of $K\inv$, where
$K\equiv\sqrt{A\tilde{f}-\tilde{\tau}^2/4}$.

An important property of $\Hop$ is that it commutes with both the
operators $\Jop\bft$ and $\hat{\Jop}\sfz$. The physical meaning of
this property is simply that a uniform rotation of the rod about the
constant axis $\sfbz$ changes nothing, and (by the rod's isotropy)
neither does uniform rotation of the rod about its own $\hat{t}$-axis. 

Thus the unnormalized propagator $\Psi$ obeys a differential equation
which is of Schr\"odinger type, in imaginary time. The derivatives
$\Jop _\alpha$ correspond to ${\rm i}/\hbar$ times the usual angular
momentum operators, and so on.
In the next section, we will exploit the quantum
mechanical analogy to find solutions to this equation which will in
turn allow us to determine the quantities $\langle z \rangle$ and
$\langle\Lk \rangle$.

\subsection{Solution and Results \label{Quantities}}

It is now possible to make a direct connection between the eigenvalue
problem associated to \eqn{pnHdef} and our polymer problem. 

In ordinary quantum mechanics, the
solution to the Schr\"odinger equation for a symmetric top can be
written as a superposition of Wigner functions:\cite{LandauQM}
\beq
\Psi(\Mg,t)=\sum_{ jmk}\ c_{ jmk} {\e}^{-{\rm i}{\cal E}_{ jmk} t}\
{\cal D}^{ j}_{ mk}(\Mg) .
\label{psisolution}
\eeq
Here $m$ and $k$ are angular momenta associated with the
operators $\Jop \bft$ and $\hat{\Jop }\sfz$,
and ${\cal E}_{ jmk}$ is the
eigenvalue associated with the Wigner function ${\cal D}^{ j}_{
mk}$.  The coefficients $c_{ jmk}$ characterize the initial state
at time $t=0$.  

It may seem difficult to apply \eqn{psisolution} to our
statistical problem, since in our case $\Jop \bft$ and $\hat{\Jop
}\sfz$ are real, antisymmetric operators with no basis of real
eigenvectors. Similarly, and unlike the case of the wormlike chain,
$\Hop$ has no particular 
symmetry. A little thought shows, however, that these are surmountable 
problems. Since one end of our rod is clamped, the initial probability
distribution $\Psi(\Mg,0)$ may be 
taken to be a delta-function concentrated on $\Mg=\ss1$, the identity
matrix $\theta=\psi+\phi=0$. This $\Psi$ is indeed an eigenstate of 
$\Jop \bft-\hat{\Jop }\sfz$ with eigenvalue $m-n=0$. The other end of
the rod may also be considered clamped to $\theta=0$, but since we
work in the fixed-torque ensemble the overall rotation $\psi+\phi$ is
free to take any value. In other words, after evolving $\Psi(\Mg,0)$
to $\Psi(\Mg,L)=\e^{-({\cal E}_0+\Hop)}\Psi(\Mg,0)$ we need to {\it
project} it to the eigenspace with $\Jop \bft+\hat{\Jop
}\sfz=0$. Since as noted earlier $\Jop \bft$ and $\hat{\Jop
}\sfz$ both commute with $\Hop$, we may perform the projection on 
$\Psi(\Mg,0)$ instead. 

Thus for our problem we should simplify \eqn{pnHdef} by setting $\Jop
\bft=0$ and $\hat{\Jop }\sfz=0$, obtaining the differential equation
that appeared in earlier 
work:\cite{BouchiatMezard,JDMPNPNAS}
$\dot{\Psi}=-(\Hop +{\cal E}_0)\Psi$, where
\beq
\Hop ={K\over A} \left[- {\vec{\Jop }^{\,2}\over{2K}} + \left(
K
-{\tilde\tau^2\over4K}{1-\cos\theta\over1+\cos\theta}
\right)(1-\cos\theta) \right] ,
\eeql{depnas}%
\beq
K\equiv\sqrt{A\tilde{f}-\tilde{\tau}^2/4}\ ,
\eeql{Kdefb}%
and ${\cal E}_0=-(\tilde{f}+\tilde{\tau}^2/2C)$.  The major difference between
this equation and that obtained for ordinary (non-twist storing) polymers is
that the long-wavelength cutoff is now controlled by $K$ instead of
$\sqrt{A\tilde{f}}$.

The operator in \eqn{depnas} really is
symmetric, and hence will have real eigenvectors (modulo a subtlety
discussed in appendix~B). The solutions
to our Schr\"odinger-like equation will then have the form
\eqn{psisolution} with i$t$ replaced by arclength $s$.
For a sufficiently long chain, the lowest ``energy''
solution will then dominate $\Psi$.  The thermodynamic properties of the
polymer can then be determined by remembering that $\Psi$, the
unnormalized propagator, becomes equal to a constant times the
partition function ${\cal Z}$, and applying \eqn{extension.link}.

We gain further confidence in the above analysis when we note that the 
terms set to zero in \eqn{pnHdef} include some which are linear in
the applied 
torque $\tau$.  For reasons outlined in section~\ref{physicalpicture},
we do not expect these terms to play a role in the determination of
the lowest energy eigenvalue.  The model that we defined is non-chiral
and therefore cannot tell the difference between over- and
undertwisting.  

We must now compute
the lowest eigenvalue of the differential operator in
\eqn{depnas}.  Finding it would be a straightforward task were it not for the
singularity in the potential term when $\theta \rightarrow \pi$.  This
singularity is associated with the backbone tangent $\hat{t}$ looping around to
point anti-parallel to the end-to-end displacement vector $+\sfbz$.
Physically, this situation corresponds to the onset of supercoiling.  When
the applied torque is too high or the tension is too low, the chain will begin
to
loop over itself.  Since real chains cannot pass through themselves, they
begin to form plectonemes.
In our {\sl phantom chain} model, there is no self-avoidance, and so
the chains can pass through themselves, shedding a unit of
$\Lk$ as they do.  The mathematical pathology associated with the
$\theta\rightarrow\pi$ singularity in \eqn{depnas} is therefore
an inevitable consequence of our model's neglect of self-avoidance.

The physical breakdown of the {\sl phantom chain} model and the
corresponding mathematical problem of the $\theta
\rightarrow\pi$ singularity can be avoided by assuming that
the backbone tangent $\hat{t}$ remains nearly parallel to the
$+\sfbz$-axis.  Such a situation is indeed realistic for a chain under
sufficient tension, or more precisely, for a sufficiently large $K$
\eqn{Kdefb}. 
In this regime, we can then perform a perturbative expansion about
$\theta=0$.  The singularity of \eqn{depnas} does not affect low
orders of perturbation theory.  The singularity can still enter
nonperturbatively via ``tunneling'' processes, in which the straight
$\theta\approx 0$ configuration hops over the potential barrier in
\eqn{depnas}, but 
these will be exponentially suppressed if the barrier is sufficiently
high, a condition made more 
precise in appendix~B.  The perturbative regime is experimentally
accessible: we will argue that it corresponds to the solid symbols on
Figure~\ref{figure1}.  Outside this regime, the {\sl phantom chain}
model is physically inappropriate, as explained above, and so a full
nonperturbative  solution of our model would not be meaningful.

We can simplify the problem by changing variables from $\theta$ to
$\rho^2\equiv2(1-\cos\theta)$. In terms of $\rho$ the spherical
Laplacian $\Jop^2={1\over\sin\theta}{\partial\over\partial\theta}
\sin\theta{\partial\over\partial\theta}$ becomes 
$(1-\rho^2/4){\partial_\rho}^2+(1-3\rho^3/4)\rho\inv\partial_\rho$, so
\beq
\Hop ={K\over A} \left[ -{{\nabla^2}\over{2K}}+{K\over 2}\rho^2
+{1\over{2K}}\left({{3\rho}\over 4}{\partial\over{\partial\rho}} +
{{\rho^2}\over4}{{\partial^2}\over{\partial\rho^2}}
-{{\tilde{\tau}^2\rho^4}\over 16-4\rho^2}\right)
\right]
\eeql{Hamiltonian}%
where $\nabla^2=\rho^{-1} \partial_\rho \rho \partial_\rho$. We have
not made any approximation yet.

We now construct a perturbative solution to the eigenvalue problem
defined by  \eqn{Hamiltonian}.
In the quantum mechanical analogy this
equation describes a two-dimensional anharmonic oscillator, with $\rho$
interpreted as a radial coordinate; thus the problem can be solved
using the method of raising and lowering operators.  

Switching to Cartesian coordinates, we set
\beq
\Aop _{\pm}=\sqrt{K \over 2}\left(x\mp{1\over K}{\partial \over{\partial
x}}\right),\qquad {\rm and}\qquad \Bop _{\pm}=\sqrt{K \over
2}\left(y\mp{1\over K}{\partial \over{\partial y}}\right).
\eeq
Now \eqn{Hamiltonian} can be rewritten as $\Hop =\Hop _0 + \delta\Hop $,
where
\bea
\Hop _0&=&{K\over A}\left( \Nop _{\rm a} + \Nop _{\rm b} + 1 \right),
\qquad {\rm and} \nonumber \\
\delta\Hop &=&{K\over A}\left[-{1\over{8 K}}
\left(1-{1\over 4} \left\{
(\Aop _+^2-\Aop _-^2)+(\Bop _+^2-\Bop _-^2)\right\}^2\right) +{\cal
O}(K^{-3})\right].
\eeal{hop0deltahop}%
Here $\Nop_{a}\equiv\Aop_+\Aop_-$ and $\Nop_{b}\equiv\Bop_+\Bop_-$
correspond to the usual occupation number operators in the quantum
mechanical analogy.  It is now straightforward to calculate the lowest
energy eigenvalue as an expansion in $K^{-1}$ to obtain\cite{JDMPNPNAS}
\beq
{\cal E}={\cal E}_0 + {K\over A} \left( 1-{1\over{4K}}-{1\over{64 K^2}} +
\cdots\right) .
\eeql{lowesteigenvalue}%
Remarkably, this is exactly the same formula as the one appearing in
the wormlike chain model; the only difference is that $K$ is now
defined by \eqn{Kdefb} instead of by $\sqrt{A\tilde f}$.
The last two terms retained will now give anharmonic corrections
to the simple lowest-order calculation announced
earlier.\cite{PhilDIMACS} The ellipsis represents terms of higher
order in $K\inv$ than the ones kept. We explore the status of such
terms in appendix~B. In particular, the last term of \eqn{Hamiltonian} 
has been dropped altogether. Since the expectation value of this term
is obviously divergent at $\rho=4$ ({\it i.e.} the antipode $\hat
t=- \hat z$), a certain amount of justification will be needed for
dropping it.

{}From this eigenvalue, the mean extension and the average linking number for a
given tension and torque can be found using \eqn{extension.link},
\eqn{psisolution}, and \eqn{lowesteigenvalue}:
\beq
\left\langle{z\over L}\right\rangle=1-{1\over{2K}}\left(1+{1\over{64 K^2}}
+{\cal O}(K^{-3})\right
), \qquad {\rm and}
\eeql{zavg}
\beq
\left\langle{{\Lk}\over{L}}\right\rangle={{\tilde{\tau}}\over{2\pi}}
\left({1\over
C}+{1\over{4AK}} +{\cal O}(K^{-3})\right) .
\eeql{Lkavg}
More accurate versions of these formul\ae{} are given in appendix~B.
By solving the second of these equations for the torque we obtain the new,
effective twist rigidity $C\eff$ by noting that
$\tilde{\tau}(\tilde{f},\Lk)\approx {{(2\pi\Lk/L)} C\eff(\tilde{f})}
+{\cal O}(K^{-3}) $, where $C\eff(\tilde{f})$ is given by the formula
\eqn{Ceff}.
This formula describes the ``thermal softening'' of the twist rigidity
alluded to earlier.  The effective rigidity $C\eff(f)$ is reduced from
the bare, microscopic value by a factor which arises from thermal
fluctuations.

Combining \eqn{zavg} and \eqn{Lkavg} together with the
definition of $K$ in \eqn{Kdefb} produces a formula for
the average end-to-end extension for a polymer subject to a linking number
constraint and an applied tension.  In section~\ref{fitresults} we will
compare this theoretical prediction to the experimental results of Strick {\it
et al.}\cite{Strick}
and Allemand and Croquette.\cite{Allemand98}

\subsection{Onset of Non-Perturbative Corrections\label{nonpert}}

The theory described above is only valid in the regime where the
{\sl phantom chain} model is appropriate.  In this section, we extend
our analysis by {\it estimating}  the effect of plectoneme formation
close to its onset.   As discussed
above, our model is unable to include such effects quantitatively, as it lacks
the
self-avoidance interaction which stabilizes plectonemes. Instead, in
this section we will suppose that the main
consequence of the singularity is to allow each segment of the polymer to be
in one of two configurations.  In the first instance, the polymer fluctuates
about
a nearly straight conformation and can therefore be described by the theory
developed in the preceding sections.  In the second instance, the polymer is
driven across the ``tunneling'' barrier into a standard kink conformation as
depicted in Figure~\ref{figure2}, gaining approximately one unit of
Writhe. Our improved formul\ae{} will have {\it no new fitting
parameters} beyond the ones already introduced.

We are interested only in the initial stages of plectoneme formation and so it
will be sufficient to approximate each plectonemic coil by a circle.  The
energy
required to form such a loop is
\beq
{{\Delta E}\over \kt}={A\over 2} {{2 \pi R} \over {R^2}}
+ \tilde{f} 2 \pi R - 2 \pi |\tilde{\tau}|.
\eeq
Here the first two terms represent the energy costs associated with bending the
polymer and contracting against the imposed tension.  The last term
gives the elastic twist energy
released as Twist gives way to Writhe.  Maximizing the energy release, we find
the optimal radius of a coiled segment to be $R=\sqrt{A/2\tilde{f}}$,
so that for $\tilde\tau>0$ the
presence of a kink lowers the energy of the polymer by
$\Delta E_-/\kt = 2\pi (\sqrt{2 A \tilde{f}}-\tilde{\tau})$.

We now imagine the polymer to be made up of segments of length $2
\pi R$.  Each of these segments may be in the extended or the
plectonemic kink configuration.  Actually, we will consider two
possible types of kink: one in which a unit of Twist is shifted into
Writhe, and its mirror image which generates a negative Writhe as well as a
counteracting positive Twist.  The reverse kinks are energetically unfavorable
for appreciable applied torque, but we retain them to eliminate any
asymmetry in the excess linking number.

For the modest applied torque considered here, it will be sufficient to treat
a dilute gas of positive and negative kinks.  Denoting the population of kinks
by $n_-$ and that of reverse kinks by $n_+$, we have
\beq
\langle n_{\pm} \rangle = {{\kappa L}\over{2\pi R}} \exp\left(-{{\Delta
E_\pm}\over\kt}\right).
\eeq
Here $\Delta E_+ = 2\pi (\sqrt{2 A \tilde{f}}+\tilde{\tau})$ is the energy of
a reverse kink, and $\kappa$ is
a numerical factor of order unity arising ultimately from a functional
determinant. Since 
we do not know how to compute $\kappa$ we set it equal to unity.

The effect of the kink/anti-kink gas is to modify \eqn{zavg} and \eqn{Lkavg},
producing shifts in the average extension and average linking number:
\beq
\Delta\left\langle{z\over L}\right\rangle = -\left(\e^{ -\Delta
E_-/\kt} + \e^{ -\Delta E_+/\kt}\right)
\eeql{pnkinka}%
\beq
\Delta\left\langle{{\Lk}\over{L}}\right\rangle =
{1\over{2\pi}}\sqrt{{2\tilde{f}}\over A}
\left(\e^{ -\Delta E_-/\kt}-\e^{ -\Delta E_+/\kt}\right).
\eeql{pnkinkb}%
These expressions are to be added to \eqn{zavg} and \eqn{Lkavg}. 
The latter
expression can then be solved for $\tilde\tau$ to get a
corrected version of \eqn{Ceff}.

The model proposed here for plectoneme formation is too simplified to give
quantitative predictions about the non-perturbative regime.  However,
the model does allow us to predict the onset of these effects and
confirm that the data we select are not affected by plectoneme formation 
(see Figure \ref{figure1}).

\section{Fit Strategy and Results \label{fitresults}}

\begin{figure}
\centerline{\epsfig{figure=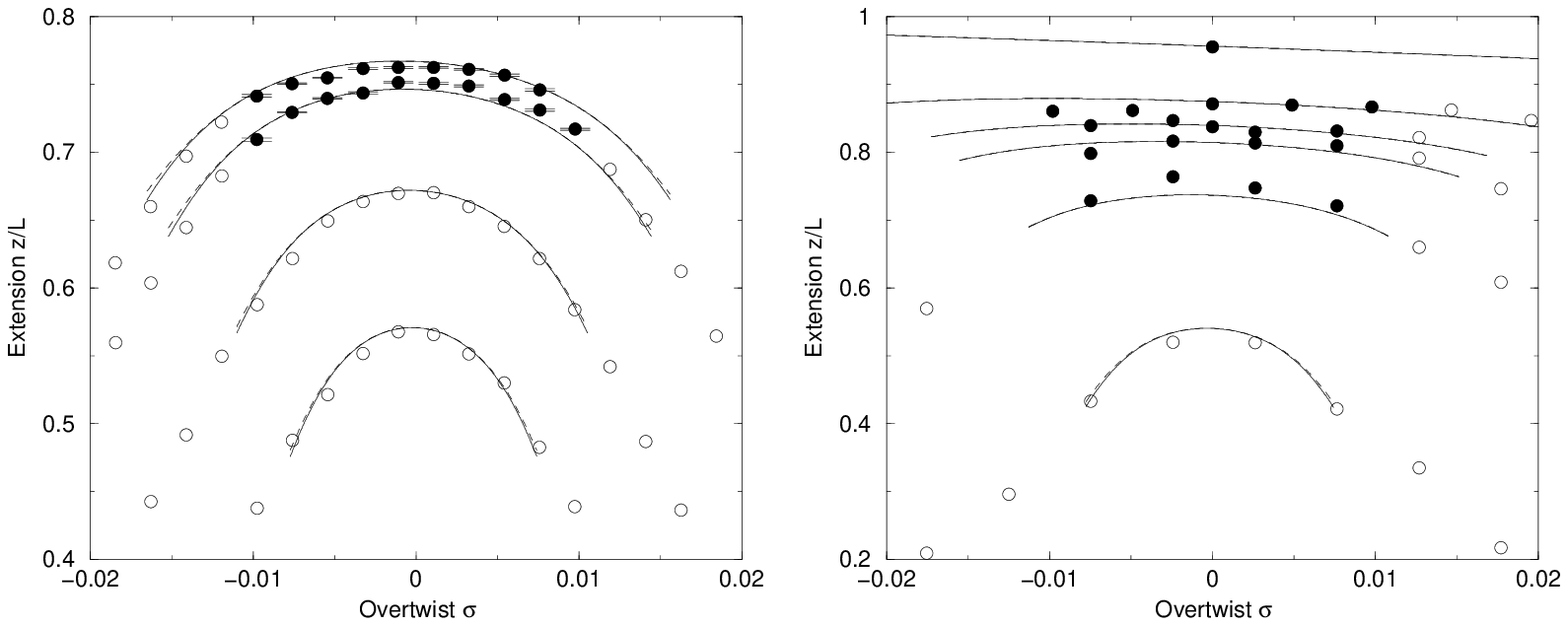,width=6.5in}}
\vspace*{0.75in}
\caption{\label{figure1}Relative extension of $\lambda$-DNA versus
applied force $f$ and overtwist $\sigma$: a single global fit to two
experiments. Fitting our model to the solid points shown correctly
predicts many of the open symbols shown, even though they were not
used in the fit. On the left are experimental data from Allemand and
Croquette:\cite{Allemand98} from top to bottom, the curves are at
fixed force 0.388, 0.328, 0.197, and 0.116~pN.  The error bars reflect
the measurement of extension; estimated errors in the determination of
the force are not shown. On the right are data from Strick {\it et
al.}:\cite{Strick} from top to bottom, the curves are at fixed force
8.0, 1.3, 0.8, 0.6, 0.3, and 0.1~pN (error estimates not
available). Points corresponding to $f,\sigma$ where the DNA is known
to denature or undergo structural change have been omitted from the
right hand graph.  Solid symbols are within the range of validity of
our model (for example, all solid symbols have $K^2>\thresh$, see
text); open symbols were not included in the fit.  A total of \Nvalue\
experimental data points were used in the fitting procedure. Some of
these points are not shown; they had force not equal to one of the ten
values listed above.  The solid lines are a single global fit to both
datasets using the theory developed in the text (see \eqn{finalz}).
The dashed (higher) lines are the same theoretical curves but without
our estimated non-perturbative contribution (section~\ref{nonpert}).}
\end{figure}

\begin{figure}
\centerline{\epsfig{figure=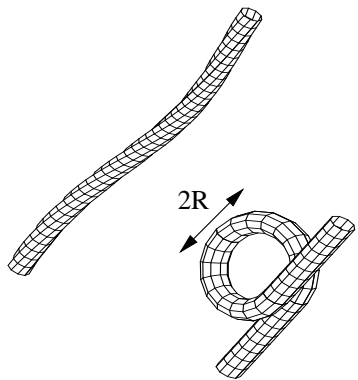,width=2.7in}}
\vspace*{0.75in}
\caption{\label{figure2}Diagram showing the idealized circular 
loop model of a plectoneme.  The twisted and slightly writhed 
conformation above is shortened by the coil circumference as 
the plectoneme forms.}
\end{figure}

The extension function $\langle z(f,\Lk) \rangle$ derived in the previous
sections describes an achiral elastic rod.  Before making direct
comparisons of this formula to experimental data, we will extend the
model somewhat.  So far we have neglected structural changes in
the DNA at a microscopic level.  In particular, we have omitted
effects related to the intrinsic stretching along the polymer
backbone.  Recent experiments have investigated these
effects;\cite{Block,Bustamante}  in particular,
Wang {\it et al.} found a small change in the relative extension of $f/\gamma$,
where
$\gamma=1100$ pN is the intrinsic stretch modulus. For moderate forces
we may simply add this shift to the extension formula found in the
previous section.\cite{Odijk,MarkoSiggiaMacro}  For the highest forces
we consider ($8.0$ pN), this translates into a relative extension of
about $0.007$, which is hardly noticeable in Figure~\ref{figure1}.
Nevertheless, we will include this correction as it improves the quality
of our fit slightly without introducing a new fitting parameter.

In addition, we will also consider the possibility of elastic
couplings which do not respect the inversion symmetry of the model that
we consider.  In reality the DNA we seek to describe is chiral, and so
at some level we expect this fact to show up as an asymmetry between
overtwisting and undertwisting in Figure~\ref{figure1}.  One
way that chirality might enter a model for DNA is through an
intrinsic twist-stretch coupling.\cite{MarkoTwistStretch,KLNO1,KLNO2}
This coupling results in a change in relative extension of $-\kt
D\omega_0^2 \sigma/\gamma$, where $D$ is the twist-stretch coefficient.  The
near symmetry in the data of Figure~\ref{figure1} indicates that
the effects of such a coupling will be small in the region of
interest.  Although the coefficient $D$ will turn out to be comparable
in size to the bending coefficient $A$, the shortening due to bend
fluctuations dominates that due to the elastic twist-stretch coupling.
This disparity arises because bend fluctuations are diverging as
$K \rightarrow 0$.  

As mentioned
in section~\ref{physicalpicture}, an anisotropy between the ``tilt''
and ``roll'' elastic constants coupled together with the associated
twist-bend coupling term might also produce an asymmetry between
positive and negative $\sigma$.  We
investigate this possibility in appendix~A and find that the
corresponding {\sl chiral entropic elasticity} terms are not
measurably different from the twist-stretch model over the range of
stretching forces studied.

Putting the intrinsic corrections associated with $\gamma$ and $D$
together with the perturbation theory result of the last section, we
obtain a theoretical prediction for the relative extension as a
function of applied force and overtwisting.  For the purposes of
comparison to experiment, we will now switch from the variable $\Lk$
to the relative overtwist $\sigma$ which is defined with respect to
the helical pitch of DNA: $\sigma=2\pi\Lk/\omega_0 L$.  Then,
\beq
\left\langle{{z(f,\sigma)}\over L}\right\rangle = 1 -
\left({2\sqrt{{{Af}\over\kt}-{{\tilde{\tau}^2}\over 4}
-{1\over
32}}}\right)^{-1}
+\Delta\left\langle{z\over L}\right\rangle
+ {{f-\kt D\omega_0^2\sigma}\over\gamma} +
{{A}\over{K^2 L}}.
\eeql{finalz}%
Formula \eqn{finalz} is our final result for the high-force (or more
precisely, large $K$) extension of a twist-storing
polymer subject to a torsional constraint. Here
$\Delta\langle{z/L}\rangle $ is the expression in \eqn{pnkinka}.
To compare our result to
the experimental data,\cite{Strick,Allemand98} we solved
\eqn{Kdefb},  \eqn{Lkavg}, and \eqn{pnkinkb} for
$\tilde{\tau}$ in terms of $\tilde{f},\sigma$, then substituted
$\tilde{\tau}$ and $K$ into \eqn{finalz}.

Apart from the intrinsic stretch and twist-stretch terms described
above, \eqn{finalz} contains two additional small refinements.  One of
these appears in the last term, where finite-size effects
have been accounted for.  This term can be understood by writing the
extension as an expansion in terms of the transverse components of the
backbone tangent. Defining the complex variable 
$\alpha(s)=\hat{t}(s)\cdot(\sfbx+{\rm i}\sfby)$ and its Fourier
components $\alpha\p$, we have to lowest order
\beq
\left\langle{z\over L}\right\rangle = 1 - {1\over 2} \sum\p \langle
\vert \alpha\p \vert^2\rangle + \cdots\ .
\eeql{tangentpert}%
The leading
entropic reduction of $\langle z\rangle$ in \eqn{zavg} is then easy to 
evaluate, including finite-length effects.
As the main
effect of an applied torque is to decrease the effective force and change the
low wavenumber cutoff in our theory from $\sqrt{A\tilde{f}}$ to $K$, we
know how to modify the usual tangent-tangent correlation function to yield
\bea
\sum\p \langle \vert \alpha\p \vert^2\rangle &=& {{4 A}\over L}\sum_{
n=1}^{\infty} {1\over{A^2 \left({{2\pi n}\over L}\right)^2
+K^2}}\nonumber \\
&=&{1\over {K}}\left( 1- {{2 A}\over{K L}}\right) .
\eeal{oneloop}%
This expression should be compared with the leading-order correction obtained
from the infinite-rod calculation in section~\ref{maincalc}:
\bea
\sum\p \langle \vert \alpha\p\vert^2\rangle &\approx& {{2
A}\over\pi}\int_0^\infty
{1\over{(A q)^2 + K^2}} \dd q \nonumber \\
&\approx&{1\over {K}} .
\eea
The difference between the two terms is $2 A / L K^2$.  To obtain the
finite length formula we must subtract this difference from the
result obtained in the last section; the resulting correction appears in
the last term of \eqn{finalz}.  Note that for the restricted values
of $K$ that we consider (see below), this contribution to $z/L$
never exceeds 0.002 for the data set we analyze.

The other refinement introduced in \eqn{finalz} is that for
convenience we replaced $({1/2K})\bigl( 
1+{1/64K^2}\bigr)$ by $\bigl(K^2-{1/32}\bigr)^{-1/2}$. Since
we will restrict our fit to $K^2>\thresh$, the difference between these
expressions is negligible. Finally, in Appendix B we give even more 
elaborate versions of \eqn{Lkavg} and \eqn{finalz}, in which
higher-order terms of perturbation theory have been retained; these
corrections are small though not negligible at low forces. 

We have now established an expression for the mean extension as a
function of applied tension and torque. Using the ENS group's
data,\cite{Strick,Allemand98} we fit this formula (actually, the more
accurate one given in appendix~B) to determine the parameters in our
model: the microscopic bend persistence length $A$, twist persistence
length $C$, twist-stretch coupling $D$, and polymer arclength $L$.  Of
these parameters, only $C$ and $D$ are really unknown; $A$ has already
been measured in other experiments, and $L$ can be determined from the
points with $\sigma=0$ using the ordinary worm-like chain model.  The
agreement between our best fit value of $A$ and earlier
experiments\cite{Block,Smith,cluzel} serves as a check on the theory.
Other parameters appearing in
\eqn{finalz}, namely $\omega_0=1.85\, {\rm nm}^{-1}$ and
$\gamma=1100\,$pN,\cite{Block} are independently known and are not
fit.

The least squares fit was performed using a gradient descent
algorithm\cite{NR} in the parameter space defined by $A$, $C$, $D$,
and $L$.  The best fit was obtained for $A=\Avalue$ nm, $C=\Cvalue$
nm, $D=\Dvalue$ nm, and $L=\Lvalue \mu{\rm m}$.  Here $L$ is the
length of the construct from Allemand and Croquette's
experiment.\cite{Allemand98} The corresponding length for the Strick
{\it et al.} data set\cite{Strick} was determined separately using the
$\sigma=0$ points from that set and was not fit. In all, $\Nvalue$
data points from the experiments of Strick {\it et al.}\cite{Strick}
and of Allemand and Croquette\cite{Allemand98} were used in the
procedure.  The data points were selected based on three criteria.
The first cuts were made on physical grounds.  It is known that for
high applied forces ($f>0.4$ pN) DNA undergoes structural
transformation or strand separation when $\sigma<-0.01$ or
$\sigma>0.03$ (D. Bensimon, private communication); here of course we
cannot use linear elasticity theory.  We therefore omitted such points
from the right side of Figure~\ref{figure1}. (No points were omitted
from the left side.) To avoid biasing the data, in the fit we excluded
the {\it symmetric} region $|\sigma|>0.01$ from the set of points used
with $f>0.4\,$pN.

The second set of cuts was applied for mathematical reasons. Our
perturbative expansion is in powers of $K^{-1}$: we required
$K^2>{\thresh}$.  We discuss this choice in appendix~B; for now we
note that perturbation theory produces excellent agreement with
experiment for the wormlike chain\cite{MarkoSiggiaMacro} even for
$K>1$.  Choosing $K^2>{\thresh}$ eliminates all of the $f=0.1$ and
0.2~pN data points from our fit.  To confirm that we were being
selective enough, we tried other values of the threshold (between 2.5
and 4.5).  This action did not significantly alter our fit results: in
every case we found $C>100\,$nm.

Finally, in addition to these two sets of data cuts, we also imposed a
``tunneling'' criterion described in appendix~B: the idea is to ensure
that the lowest energy eigenvalue of the operator $\Hop_0$ in
\eqn{hop0deltahop} is smaller than the barrier that restrains the
system from falling into the unphysical singularity.

The reasonable agreement in Figure~\ref{figure1} between our
theoretical curves and the data {\it outside the region we fit}
(including the $0.1$ and 0.2~pN curves) indicates that our
choice of cuts is a conservative one.  As a further check, the dashed
lines in Figure~\ref{figure1} show our fitting function without the
non-perturbative correction described in section~\ref{nonpert}: we see
that these lines do not deviate from the solid lines in the
range of data we retained.

\section{Discussion}

The global fit shown in Figure~\ref{figure1} indeed resembles the 
experimental data. 
The least squares fit determined the bending stiffness, the twist
rigidity and the intrinsic twist-stretch coefficient of DNA.  As
stated earlier, the fit to the bending rigidity produced the known
value and thus serves as a check on the theory.  The chiral asymmetry
is a small effect, and so the available data do not afford a precise
determination of the twist-stretch coupling $D$. Thus our fit is
mainly a measurement of $C$.

The twist rigidity obtained by the fitting procedure is somewhat
higher than what earlier experiments have found (see section
\ref{physicalpicture}). We cannot give a quantitative estimate of our
fit parameter errors, since some of the data\cite{Strick} do not have
error bars, but we note that forcing $C=\badC\,$nm or less gives a
visibly bad fit. One might worry that this discrepancy was due to some
sort of failure of perturbation theory, despite our great care on this
point. The fact that we keep finding large $C$ as we tighten the data
cuts gives us additional confidence on this point.  Similarly, our
large value is not an artifact of DNA denaturation induced by tension,
since that would lead to a spuriously {\it low} fit value. There
remains the intriguing possibility that on the contrary, imposed
tension {\it suppresses} spontaneous local denaturation, increasing
the integrity of the DNA duplex (J. M. Schurr, private communication);
in this case our large $C$ more accurately reflects the linear
elasticity than the other, lower, values.

The discrepancy with earlier work may be more apparent than real,
however: if we do not allow for a tension-dependent thermal reduction
of the twist rigidity as in \eqn{Ceff} and instead fit the data to a
constant twist rigidity, then we obtain $C\eff=\Ceffvalue$ nm, a value
closer to those found in the other experiments.\cite{HeathJMB} The
quality of this fit, obtained with a tension-independent rigidity, is
slightly poorer.  In any case, a large value of $C/A$ is not
paradoxical and in particular need not imply a negative Poisson ratio
for our model's rod: random natural bends in DNA reduce the effective
bend stiffness $A$ measured in stretching experiments, but not
$C$,\cite{NelsonDisorder} and so the ratio of $C$ to the true elastic
bend stiffness is closer to unity than it appears from our
effective-homopolymer model.

Recently, Bouchiat and M\'ezard\cite{BouchiatMezard} have also
determined the twist rigidity of DNA using the experimental results of
Strick {\it et al.}\cite{Strick} They derived formul\ae\ equivalent to
\eqn{zavg} and
\eqn{Lkavg}.  Then using an exact ground state solution to a cut-off
version of \eqn{Hamiltonian}, they reproduced the observed extension
curve $\langle z(f,\sigma)\rangle$ in Figure~\ref{figure1}a over a
wider range than we have shown.  The result of this calculation is a
ratio of $C/A$ of approximately $1.7$.

While both approaches are similar, our perturbative approach precludes
us from analyzing the lowest force curves that Bouchiat and M\'ezard
discussed.  As described above, we excluded these data because we
expect physical difficulties with the {\sl phantom chain} model in
this regime; the same difficulties, it would seem, apply to the
analytical results of Bouchiat and M\'ezard.  In particular, at small
applied tension, the backbone's tangent vector $\hat{t}$ will wander
from the $z$-axis.  If it wanders too far, the system will be able to
see through the tunneling barrier to the singularity; or in other
words, the results will be corrupted by the failure of Fuller's
formula for $\Wr$.  Bouchiat and M\'ezard approached this problem by
introducing a new intermediate-length cutoff $b=6\,$nm into the
problem. The physical meaning of this cutoff in terms of the
mechanical properties of DNA is not clear to us. Moreover, taking it
to be 2.5~nm or less spoiled the simultaneous fit at all values of
$f$.

In contrast, our perturbative treatment avoids the singular-potential
problem altogether by restricting to a regime where the {\sl phantom
chain} model is valid.  Our model has no extra scale corresponding to
$b$, and yet fits all fixed-force curves in its domain, in two
different experiments, with one value of $C$.

In their paper, Bouchiat and M\'ezard also gave Monte Carlo results.
Earlier work by Marko and Vologodskii has also taken this
approach.\cite{MarkoVolo} Here it is possible to implement
self-avoidance, though knot rejection is still difficult.  The
advantage of analytic formul\ae\ such as \eqn{finalz} is that they
permit global, systematic least-squares fitting of $\langle
z(f,\sigma)\rangle$ to the data. Moreover, for practical reasons Monte
Carlo simulations must again impose a short-distance cutoff of at
least several times the DNA radius, unlike our analytical approach.

\section{Conclusion}

In this paper we have investigated the statistical mechanics of a
twist-storing polymer.  This type of molecule differs from a
traditional polymer in being unable to relax out an applied excess
Link.  When such a chain is left unconstrained, the twist simply
decouples from the bend fluctuations.  The thermally accessible
conformations are then identical to those for an ordinary polymer.  In
the case that such a polymer is subject to a torsional constraint,
however, there will be a coupling between the bend fluctuations and
the twist.  It is this coupling that we have investigated. One of our
goals was to show how single-molecule stretching experiments can
provide a new window onto the nanometer-scale mechanical properties of
DNA.

Due to the complications associated with self-avoidance, we considered
only chains held nearly straight by tension, then analyzed the
statistical mechanics of the resulting ``torsional directed walk''.
We mapped the polymer partition function onto the solution of a
Schr\"odinger-type equation for the orientation distribution function.
From this solution, we were able to find the entropic extension and
the overtwisting of a polymer subject to a tension $f$ and relative
Link excess $\sigma$.

The theory we developed quantitatively reproduces the results of
supercoiled single-molecule DNA stretching
experiments\cite{Strick,Allemand98} (see Figure~\ref{figure1}).  The
agreement was achieved by fitting the twist persistence length,
yielding $C=\Cvalue$ nm.  The large twist rigidity differentiates DNA
from traditional polymers and makes possible the coupling of the twist
and bend degrees of freedom that plays a central role in our theory.

Apart from reproducing the experimentally observed physics, our
formul\ae\ make another prediction: the twist rigidity is renormalized
(see \eqn{Lkavg}).  The effective rigidity $C\eff(f)$ is a function of
the applied tension.  According to \eqn{Ceff}, it is hardest to twist
the polymer when it is pulled straight; this is the bare, microscopic
stiffness.  It is the same rigidity that resists twist at the shortest
length scales, and so enters the energetics of structures such as the
nucleosome. As the tension is relaxed, thermal fluctuations begin to
play a role.  Now when a torque is applied, the polymer does not
resist as much; the bend fluctuations have softened the torsional
rigidity by absorbing some of the imposed excess Link.  As discussed
above, this phenomenon is purely thermal; no such effect appears in
the linear elasticity of a macroscopic beam for small applied torque.

If one na\"{\i}vely extends this thermal effect to zero tension, one
sees that the torsional rigidity vanishes completely.  Of course, our
{\sl phantom chain} model precludes us from considering this case;
however, other recent work\cite{MorozKamien} has considered this
related problem using an explicit self-avoidance term: indeed the
effects of a torsional rigidity do become unimportant to the behavior
of twist-storing polymers at zero applied tension or, equivalently, at
extremely long length scales.

\section*{Appendix A: Chiral entropic elasticity\label{appa}}

In this appendix we introduce an additional element of realism into
our model, namely the intrinsic helical pitch $2\pi/\omega_0$. For DNA
this pitch corresponds to $\omega_0=1.85/$nm. The helical structure
breaks the inversion symmetry of the problem by allowing two
additional terms in the energy functional.\cite{MarkoSiggiaBendTwist}
In principle these explicitly chiral terms could introduce an
asymmetry between overtwist and undertwist into our results. We will
find this {\sl chiral entropic elasticity} and show that it has a
different dependence on stretching force from the intrinsic
twist-stretch effect discussed in section~\ref{fitresults}. Thus in
principle the two effects could be distinguished experimentally.

In this appendix we are interested in chiral effects, manifested by
odd powers of $\sigma$ in the extension $z(f,\sigma)$, in a model of
DNA without intrinsic stretching. We will see that such terms are
small. Hence we can use a simpler calculation than the one in the main
text: we will drop ${\cal O}(\sigma^2)$ and higher, and we will use
the Gaussian (or equipartition) approximation to the statistical
sums. Since odd-power terms are completely absent in the achiral model
of section~\ref{maincalc} above, we can simply add the ones we find to
the results of that model to get a leading approximation to the full
{\sl chiral entropic elasticity} formula.

Another approximation we will make will be to drop terms suppressed by
powers of $1/{\omega_0}$, since this length scale is much shorter than
both the persistence lengths and the scale $\sqrt{A/\tilde{f}}$ of
important fluctuations.

As discussed in section~\ref{physicalpicture}, chirality can enter
through the anisotropic bending rigidities associated with the
``roll'' and ``tilt'' axes of DNA monomers. In this appendix we will
choose a material frame different from the one in the main text: here
our frame rotates with the intrinsic helical twist. This choice is
convenient in that the anisotropic elasticity appears constant in this
frame: \eqn{bendtwist} becomes simply
\beq
{{E_{\rm bend}}\over\kt}={1\over 2} \int_0^L \dd s \left( A'\bfn
{\Omega\bfn}^2 + A'\bfm{\Omega\bfm}^2\right) ,\qquad{\rm and}\qquad
{{E_{\rm twist}}\over\kt}={1\over 2}\int_0^{ L}\ C'{\Omega\bft}^2 \
\dd s .
\eeq
Here we have introduced {\it two} microscopic bending constants,
$A'\bfn$ and $A'\bfm$.  Now even the unstressed state will be chiral:
as the body-fixed frame $\{\bfbx,\bfby,\bfbz=\hat t\,\}$ rotates
uniformly at frequency $\omega_0$ along the polymer, it turns the bend
anisotropy with it. Since the $\bfbx$-axis corresponds to the short
axis of a basepair, we expect $A'\bfn>A'\bfm$.

Apart from the bending anisotropy, the symmetries of DNA admit an
explicitly chiral term associated to a twist-bend coupling with
coefficient $G$.\cite{MarkoSiggiaBendTwist} With these two terms, the
mechanical-equilibrium state of the stressed molecule will no longer
be given by the uniformly twisted configuration.  Instead, we make an
{\it ansatz} for a new helical ground state: $\Mg _0=\exp(\zeta \ML_1)
\exp(\omega s \ML_3)$, to be justified below.  Here $\omega$ includes
a finite piece associated with the rotation of the unstressed
molecule, so that $\omega=\omega_0 (1+\sigma)$.  The small angle
$\zeta$ remains to be determined by the condition of mechanical
equilibrium.  The elastic energy functional for the model is then
given by:
\beq
{E\over\kt}={1\over 2}\int_0^{ L} \dd s \left\{A'\bfn {\Omega\bfn}^2 +
A'\bfm {\Omega\bfm}^2 + C' ({\Omega\bft}-\omega_0)^2+ 2 G {\Omega\bfm}
({\Omega\bft} - \omega_0)\right\} - \tilde{f} \cdot z.
\eeql{pnchelas}%
In contrast to the discussion in the main text, in this appendix we
will work in the fixed-$\Lk$ ensemble. Thus we do not need any
Lagrange multiplier associated with the Link constraint.

It will prove convenient to introduce the combinations
$\bar{A}=(A'\bfn+A'\bfm)/2$ and $\hat{A}=(A'\bfn-A'\bfm)/2$. We
emphasize that $\bar{A}$ is not necessarily equal to $A$ from the
coarse-grained model \eqn{bendtwist}; the exact relationship will
emerge in due course below.  The chiral terms that couple to the
intrinsic helical frequency $\omega_0$ are then proportional to
$\hat{A}$ and $G$.  Note that $\hat{A}>0$.

We can now determine the helix angle $\zeta$ characterizing the
mechanical-equilibrium state. First write a small fluctuation from
$\Mg_0$ as $\Mg(s)=\tilde{\Mg}(s)\Mg_0(s)$ with $\Mg_0$ as above and
$\tilde{\Mg}(s)\equiv
\e^{-T_\alpha (s)\ML_\alpha}$. Substituting into (\ref{pnomdef},\ref{pnomcomp})
then yields the $\Omega_i$'s. Setting the first variation of
\eqn{pnchelas} to zero then yields
three equations expressing the condition that $\Mg _0$ be the stressed
mechanical-equilibrium state.  One of these selects $\zeta$:
\beq
\zeta=-{{G \sigma}\over{A'\bfm  + \tilde{f}/{\omega_0}^2}}
\approx-{{G \sigma}\over{A'\bfm}}
.\eeq The other two are satisfied trivially, justifying our {\it
ansatz} for $\Mg_0$. In deriving the above relations we used the fact
that we are working in the fixed-$\sigma$ ensemble. Thus the boundary
conditions clamp the rod at both ends, fixing $\vec{T}=0$ there, and
so we may discard total derivative terms.

For illustration, and to keep the calculation simple, we will now make
the additional assumption that the chiral parameters $\hat{A},G$ are
both smaller than $\bar{A},C$, and accordingly work to leading
nontrivial order in the former. We can then easily diagonalize the
part of the energy involving the latter using Fourier modes.  Setting
$\bigl(T_1(s) + {\rm i} T_2(s)\bigr)\e^{{\rm i}\omega s}\equiv
\sum {\e}^{{\rm i} q s} \alpha_q$ and
$T_3(s)\equiv\sum {\e}^{{\rm i} q s} \phi_q$ (note that
$\phi_{-q}=\phi_q^*$) yields
\beq
{E\over\kt}=-\tilde{f}\cdot L + \xi_0 +\xi_1+\xi_2 ,\eeq where
\bea
\xi_0 &=& {L\over 2} \sum\p \left[ \left(\bar{A} p^2 - \omega (
C'\sigma + 2G\zeta) p +\tilde{f}\right) \vert \alpha\p \vert^2 + C'
p^2
\vert \phi\p \vert^2 \right] \nonumber \\
\xi_1 &=& {{\rm i} L\over 2} \sum\p \left[
G(\omega-p)p +\zeta\left( A'\bfn\omega(\omega-p)+\tilde{f}-C'\omega p
\right)
\right]
\left(\phi_p\alpha_{\omega-p}-{\rm c.c.}\right) \nonumber \\
\xi_2 &=& {L\over 2} \sum\p \left[{\hat{A}\over 2} p (p - 2 \omega )
-{{G\omega \zeta}\over 2}p \right]\left( \alpha\p \alpha_{ 2\omega-p }
+ {\rm c.c.}\right).
\eea
In the above formul\ae, $\omega\equiv\omega_0 (1+\sigma)$ gives the
angular frequency for the stressed minimal-energy state. The sums are
for $-\infty<p<\infty$ (the physical short-scale cutoff will prove
immaterial). As mentioned above, we will treat $\xi_{1,2}$ as
perturbations to $\xi_0$.

In the harmonic approximation, the mean extension has the simple form
\beq
\left\langle {z\over L}\right\rangle = {1\over L} {\dd\over{\dd\tilde{f}}} \ln
{\cal Z} = 1 - {1\over 2}\sum\p\langle \vert \alpha\p \vert^2 \rangle
+ \cdots .
\eeql{pnzl}%
We define $D(p)\equiv L \langle \vert \alpha\p \vert^2 \rangle$ and
compute this two-point correlator perturbatively.

The unperturbed $D_0(p)$ is obtained via equipartition, or
equivalently by performing the Gaussian (harmonic approximation)
functional integral over $\alpha\p$ and $\alpha\p^*$ in $\xi_0$,
yielding
\beq
D_0 (p) = L \langle \vert
\alpha\p\vert^2\rangle_0={{2}\over{\bar{A}(p^2-2 q_0 p) +
\tilde{f}}},
\eeq
where
\beq
q_0 = \left( C' - {{2 G^2}\over{A'\bfm }}\right){{
\omega \sigma}\over{2 \bar{A}}}.
\eeq
The next step in determining $D(p)$ is to calculate the first two
corrections, $\Pi_1 (p)$ and $\Pi_2(p)$, induced by $\xi_1$ and
$\xi_2$ respectively. We define these as lowest-order corrections to
the full two-point function: $D(p)\equiv
D_0(p)[1+D_0(p)(\Pi_1(p)+\Pi_2(p))]$.  Start by expanding the
contribution ${\e}^{-\xi_2}$ to the Boltzmann factor.  There is no
first order correction, so we go to second order:
\bea
\Pi_2 (p) &=& 
p^2{(\hat{A}(2\omega-p)+G\omega\zeta)^2
\over\bar{A}((2\omega-p)^2-2q_0(2\omega-p)+\tilde{f}}
\nonumber \\
&\approx& p^2{\hat{A}^2+\zeta\hat{A}G\over\bar{A}} .
\eeal{D2}

The second correction arises from the expansion of the energy in
powers of $\xi_1$.  Once again we go to second order:
\beq
\Pi_1 (p) = 
{{\bigl(G^2p^2+2G\zeta(A'\bfn p^2-C'\omega p)\bigr)}\over{2C'}}.
\eeql{D1}%
As mentioned above, we have dropped terms of order $\sigma^2$ and
higher: only odd-power terms will create chiral corrections to the
extension curve, and we content ourselves with investigating the
linear ones only.  Putting the results of \eqn{D1} and \eqn{D2}
together gives the propagator
\beq
D (p) = \left[D_0(p)\inv-\Pi_1(p)-\Pi_2(p)\right]\inv.
\eeql{pnpi1}%
To get \eqn{pnpi1} we summed chains of Gaussian graphs, similarly to
the random-phase approximation in many-body theory.

The relative extension can now be computed from \eqn{pnzl}:
\bea
\left\langle {z\over L}\right\rangle &=&1-{1\over2L}\sum_pD(p)\nonumber\\
&=&1-{1\over4\pi}\int_{-\infty}^\infty\dd p\,D(p)\nonumber\\
&=&1-\half\left(A\tilde{f}(1+F\sigma)\right)^{-1/2}
.
\eeal{pnzlresult}%
In this formula we have identified $A\equiv\bar{A}-2\hat{A}^2/\bar{A}
-G^2/C'$ as the {\it effective bend constant}, coarse-grained over a
helix turn. (Had we kept ${\cal O}(\sigma^2)$ terms we could have made
a similar identification of the coarse-grained twist constant $C$ in
terms of $\bar A,\hat A,C',G$.) We also defined
\beq
F\equiv {2G^2\over A'\bfm A}\left({\hat{A}\over\bar{A}}+{A'\bfn\over
C'}\right)
\eeql{pnFdef}%
The key observation is now simply that $F$ in \eqn{pnFdef} is {\it
positive}.

Thus we have found a {\sl chiral entropic elasticity} effect: the
formul\ae{} of the main text for $1-z/L$ get multiplied by the
asymmetric correction factor $(1-F\sigma/2)$. (This analysis corrects
an erroneous claim\cite{KLNO1} that no such factor exists.)

The dependence of this chiral contribution to $z/L$ on the stretching
tension is different from the intrinsic twist-stretch term introduced
in the main text, equation \eqn{finalz}, and so in principle the two
effects could be disentangled by fitting to data. In practice,
however, the chiral effect in Figure~\ref{figure1} is too small to
make any definite statement. Instead we tried eliminating the $D$ term
in \eqn{finalz} and replacing it by the $F$ term in \eqn{pnzlresult},
which yields an equally good fit but with $F=\Fvalue$.  Since this
value is not positive, contrary to the prediction in \eqn{pnFdef}, we
conclude that the twist-stretch coupling $D$ is needed to explain the
asymmetry of the experimental data. This conclusion is qualitatively
consistent with an earlier analysis\cite{KLNO1} of the highest-force
data; here the chiral entropic effect is very small (see
\eqn{pnzlresult}).  Encouragingly, the fit values of $A,C$ are similar
to those quoted in the main text --- our measurement of $C$ is not
sensitive to the precise mechanism of chiral symmetry breaking.

\section*{Appendix B: Domain of validity\label{appb}}

In this appendix we endeavor to justify our
perturbative approach to torsional directed walks, and in particular
establish its domain of validity and hence the subset of the
experimental data which falls into that domain.

\subsection*{Tunneling}

As we have mentioned several times, the Schr\"odinger-type equation
defined by \eqn{depnas} suffers from a singularity at the antipode
$\theta=\pi$. Indeed, the operator $\Hop$ has no eigenstates at all.
We have emphasized that this singularity is caused by our unphysical
omission of self-avoidance effects, but it is still necessary to have
some criterion for when the details of the nonlocal interaction
correcting the
problem will be unimportant, and some practical scheme for calculating
in this regime.\cite{lang67a,zinn84a,Colemanaspects}

The key point to note is that if we let $t\equiv\tau^2/4$ and imagine
solving our problem for {\it negative} (unphysical) values of $t$,
then our problem disappears. Analytically continuing the ground-state
eigenvalue in the complex $t$-plane back to positive (physical) $t$
yields a result which is finite but no longer real: for small $t$ its imaginary
part gives the probability of a rare {\it barrier
penetration} process. The real part is an approximate eigenvalue
describing the metastable state and controlling the intermediate
asymptotics of $\Psi$: this is the number we seek. When the imaginary
part is small, the real part can be obtained from the lowest orders of
perturbation theory, even though eventually at high orders the series
diverges. 

We can estimate the imaginary part of the eigenvalue by finding the
saddle point (or ``instanton'' or ``bounce'' or ``domain wall''
solution) of the functional integral giving rise to \eqn{depnas}.
This is the function $\theta(s)$ satisfying the ordinary differential
equation $A^2\ddot\theta=\dd V/\dd\theta$, where
$V(\theta)\equiv(1-\cos\theta)\left(K^2-t (1-\cos\theta)/(1+\cos\theta)
\right)$. The elastic energy of this configuration is then given by 
\beq
{{\bar{E}}\over\kt}=2\int_0^{\theta_1}\dd
s\Bigl[{A\over2}{\dot\theta}^2+{1\over A}V(\theta)\Bigr]
=2\int_0^{\theta_1}\dd\theta\sqrt{2V},
\eeql{action}%
where $\theta_1$ is the ``turning point'', where $V(\theta_1)=0$. The
imaginary part of the analytically-continued eigenvalue is then
proportional to $\e^{-\bar{E}/\kt}$.
Numerical evaluation shows that this factor is smaller than 0.02 when
$t<0.6(A\tilde f - 1.6)$, and we have imposed this as one of the
conditions selecting the data points used in Figure~\ref{figure1}.

\subsection*{Perturbation theory}

From the previous subsection and the references cited there we know
that when the tunneling criterion is satisfied perturbation theory
will be an asymptotic expansion, which we may approximate by its first
terms. In this subsection we will quote the eigenvalues of
\eqn{Hamiltonian} obtained using second-order perturbation theory. In
the last term we expand $\rho^4 / (1-\rho^2/4)$ in power series,
since each succeeding term is formally suppressed by a power of
$K\inv$; we keep the terms $\rho^4+\rho^6/4$. We again abbreviate
$t\equiv \tilde\tau^2/4$.

Using the operator notation of the main text, we find
$\Hop=\Hop_0+\delta\Hop$, where $\Hop_0$ is the first line of
\eqn{hop0deltahop} and
\bea
\delta\Hop &=&{1\over 8A}\biggl[
-{t\over32K^3}\bigl(\Aopp^6+3\Aopp^2\Bopp^4+3\Aopp^2\Bopp^4+\Bopp^6\bigr)\nonumber\\
&&+{1\over4}\Bigl(1-{t\over K^2}-{9t\over4K^3}\bigr)\bigl(\Aopp^4+2\Aopp^2\Bopp^2+\Bopp^4\Bigr)
\nonumber\\
&&-t\Bigl({2\over K^2}+{9\over4K^3}\Bigr)\bigl(\Aopp^2+\Bopp^2\bigr)
-2\Bigl(1+{t\over K^2}+{3t\over4K^3}\Bigr)
\biggr],
\eeal{hideous}%
plus terms annihilating the perturbative ground state. From this we
compute zeroth through second-order shift:
\bea
{\cal E}&=&-\tilde f-{2t\over C}+{K\over A}\Bigl(W_1-{t\over4K^3}W_2-\Bigl({t\over4K^3}\Bigr)^2W_3\Bigr)
{\rm \ \ where}\nonumber\\
W_1&=&1-{1\over4K}-{1\over64K^2}\nonumber\\
W_2&=&1+{5\over8K}-{9\over32K^2}\nonumber\\
W_3&=&{9\over4}\bigl(1+{5\over2K}+{16\over9K^2}\bigl)
.\eeal{disgusting}%
Taking thermodynamic derivatives as in the text (see \eqn{extension.link}),
and recalling $t\equiv\tilde\tau^2/4$, gives
\def\tts{{\tilde\tau}^2}
\bea
\omega_0\sigma&=&\tilde\tau\biggl[
{1\over C}+{1\over4AK}\Bigl(
1+{1\over2K}+{21\over64K^2}+{\tts\over16K^3}
\bigl(2+{15\over8K}+{9\over8K^2}\bigr)
+M\Bigr)
\biggr]
\nonumber\\
{z/ L}&=&1-{1\over2K}\Bigl(
1+{1\over64K^2}+{\tts\over16K^3}
\bigl(2+{15\over8K}-{9\over8K^2}\bigr)+M
\Bigr),
\eeal{yuck}%
where
\beq
M=\Bigl({\tts\over16K^3}\Bigr)^2\Bigl({9\over4}\Bigr)\Bigl(5+{15\over K}+{112\over9K^2}\Bigr)
.\eeql{Mdef}
The corrections for kinks, \eqn{pnkinka} and \eqn{pnkinkb}, and the
other corrections in \eqn{finalz} must be added to the expressions
\eqn{yuck}. The resulting formul\ae{} are the ones actually used in
the fit shown in Figure~\ref{figure1}.

We are now in a position to state the conditions for perturbation
theory to be useful. Our expansion is in powers of $K\inv$ and
${\tts/16K^3}$, so both of these must be small. To be more
precise, we imagine holding the force $\tilde f$ fixed while varying
the torque $\tilde \tau$, as in the experiment. The coefficient of
$\tts$ in $z/L$ then gives the information we need to obtain the twist
stiffness. 
Comparing the highest-order term of this coefficient retained above to
the leading term, we find their ratio to be less than 10\% when
$K^2>3$. This explains another of the cuts made on the data in the
text. We should also require that $\tts/16K^3$ be small, but this is
automatically satisfied when the other imposed conditions are.  

\section*{Acknowledgments}

We thank B.~Fain, R.D.~Kamien, T.C.~Lubensky, J.F.~Marko, C.S.~O'Hern,
J.~Rudnick, J. M.~Schurr, and M.~Zapotocky for helpful discussions, C.~Bouchiat
and M.~M\'{e}zard for correspondence, and J.-F. Allemand, D.~Bensimon, and
V.~Croquette for supplying us with experimental details and the numerical
data from references~\cite{Strick,Allemand98}.  This work was
supported in part by NSF grant DMR95-07366.  JDM was supported in 
part by an FCAR graduate fellowship from the government of Qu\'ebec.

\bigskip

\newcommand{\noopsort}[1]{#1} \newcommand{\printfirst}[2]{#1}
  \newcommand{\singleletter}[1]{#1} \newcommand{\switchargs}[2]{#2#1}

\newpage
\newpage
\end{document}